# Dichotomous Dynamics of Magnetic Monopole Fluids

Chun-Chih Hsu[1†], Hiroto Takahashi[1†], Fabian Jerzembeck[1,2], Jahnatta Dasini[3], Chaia Carroll[3], Ritika Dusad[1], Jonathan Ward[3], Catherine Dawson[3], Sudarshan Sharma[4], Graeme Luke[4], Stephen J. Blundell[1], Claudio Castelnovo[5], Jonathan N. Hallén[5,6], Roderich Moessner[6] and J.C. Séamus Davis[1,2,3,7]

1. Clarendon Laboratory, Oxford University, Parks Road, Oxford, OX1 3PU, UK
2. Max-Planck Institute for Chemical Physics of Solids, D-01187 Dresden, Germany
3. Department of Physics, University College Cork, Cork T12 R5C, Ireland.
4. Department of Physics, McMaster University, Hamilton Ontario, Canada
5. TCM group, Cavendish Laboratory, University of Cambridge, Cambridge CB3 0HE, UK
6. Max Planck Institute for the Physics of Complex Systems, 01187 Dresden, Germany
7. Department of Physics, Cornell University, Ithaca, NY 14853, USA.
†       Contributed equally to this project.

A recent advance in the study of emergent magnetic monopoles was the discovery that monopole motion is restricted to dynamical fractal trajectories (J. Hallén *et al*, *Science* 378, 1218 (2022)) thus explaining the characteristics of magnetic monopole noise spectra (Dusad, R. *et al. Nature* 571, 234 (2019); Samarakoon, A. M. *et al. Proc. Natl. Acad. Sci.* 119, e2117453119 (2022)). Here we apply this new theory to explore the dynamics of field-driven monopole currents, finding them comprised of two quite distinct transport processes: initially swift fractal rearrangements of local monopole configurations followed by conventional monopole diffusion. This theory also predicts a characteristic frequency dependence of the dissipative loss-angle for AC-field-driven currents. To explore these novel perspectives on monopole transport, we introduce simultaneous monopole current control and measurement techniques using SQUID-based monopole current sensors. For the canonical material $Dy_2Ti_2O_7$, we measure $\Phi(t)$, the time-dependence of magnetic flux threading the sample when a net monopole current $J(t) = \dot{\Phi}(t)/\mu_0$ is generated by applying an external magnetic field $B_0(t)$. These experiments find a sharp dichotomy of monopole currents, separated by their distinct relaxation time-constants before and after $t{\sim}600$ μs from monopole current initiation. Application of sinusoidal magnetic fields $B_0(t) =$ $B\cos(\omega t)$ generates oscillating monopole currents whose loss angle $\theta(f)$ exhibits a characteristic transition at frequency $f \approx 1.8$ kHz over the same temperature range. Finally, the magnetic noise power is also dichotomic, diminishing sharply after $t \sim 600$ μs. This complex phenomenology represents a new form of heterogeneous dynamics generated by the interplay of fractionalization and local spin configurational symmetry.

The existence of a fluid of emergent magnetic monopoles[1,2] in pyrochlore spin-ice materials[3] is now well attested[4]. However, a comprehensive microscopic understanding of the dynamics of monopole transport currents[5] remains a profound challenge. In such spin-ice compounds[3,4], e.g. $Dy_2Ti_2O_7$ and $Ho_2Ti_2O_7$, the lowest energy magnetic excitations are emergent magnetic charges (monopoles)[1,2]. Each $Dy^{3+}$ or $Ho^{3+}$ magnetic ion occupies a vertex of the corner-sharing tetrahedral sublattice and exhibits only two magnetic states with dipole moments $\mu \approx 10\mu_B$, pointing either towards or away from the center of each tetrahedron (Fig. 1A). Moreover, the lowest energy configuration of each tetrahedron is constrained by the dipolar-spin-ice Hamiltonian[6] to have two spins pointing in and two pointing out (2in-2out), while the higher energy excitations are the effective magnetic charges ($+m$ for 3in-1out and $-m$ for 1in-3out) that are in some degree mobile[2]. Explaining the unusual magnetization dynamics of both $Dy_2Ti_2O_7$ and $Ho_2Ti_2O_7$ has proven perplexing[7-15] and a complete microscopic transport theory of magnetic monopole currents remains an outstanding challenge. Monte Carlo simulations[16] using the standard model (SM) of spin-ice dynamics[5,17,18] as well as theoretical modelling based on random walk theory[19] indicated that thermally activated generation recombination processes and monopole motion give rise to magnetic noise with power spectral density[16,19], $S(\omega, T) \propto \tau(T)/(1 + (\omega\tau(T))^b$ versus angular frequency $\omega$, temperature $T$ and relaxation time $\tau$ with $b$ approaching 2. However, when discovered [20,21] the actual magnetic noise exhibited $S_M(\omega, T) \propto \tau(T)/(1 + (\omega\tau(T))^{b(T)})$, whose anomalous noise 'color' with exponent approaching $b(T) = 1.5$ at lowest temperatures represented an outstanding mystery.

Innovative monopole transport theories designed to address this issue now posit that the microscopics of $Dy^{3+}$ spin flips plays a central role in shaping the monopole dynamics. This is adduced to internal transverse magnetic fields being strongly suppressed at $Dy^{3+}$ ion sites for a highly symmetric local spin configuration as shown schematically in the side panels of Fig. 1A. In consequence, there are predicted to be two microscopic $Dy^{3+}$ spin-flip rates with fast and slow time constants [22]. Such route-dependent spin flip restrictions permeate the spin ice crystal and force peripatetic monopoles to traverse a disordered cluster of trajectories whose fractal structure is close to a percolation transition [23]. To capture the physics of the two distinct spin-flipping rates, the beyond standard model (bSM) dynamics is simulated by considering the transverse field from the six nearest spins. In particular, monopole hopping occurs on a slow time scale when the spin-flipping transverse field vanishes, and on a fast time scale otherwise. In practice, the slow time scale is expected to be so much longer than the fast one that monopole hopping at this timescale can be neglected (see Ref. 23 and SI *Appendix* section I). The theoretical consequence, supported by bSM Monte Carlo simulations [23] with this bimodal spin-flip constraint, is sub-diffusive monopole motion for which $S_M(\omega, T) \propto \tau(T)/(1 + (\omega\tau(T))^{b(T)})$ with $b(T) \approx 1.5$. This noise power law is strikingly consistent with the reported experimental magnetic monopole noise spectroscopy results [20,21].

**Analytical and numerical monopole transport theory**

The original research using the bSM theory [23] for monopoles in spin-ice focused on magnetic fluctuations and magnetic noise in thermodynamic equilibrium. But this paradigmatic change in understanding also holds profound but unexplored repercussions for monopole current dynamics of these systems when externally driven. Heuristically, we can consider these issues by visualizing a spin-ice sample as shown schematically in Fig. 1A. While the monopole drift velocities ($v_+$, $v_-$) from oppositely charged monopoles are in opposite directions, the monopole currents ($J_+ \propto mv_+$, $J_- \propto -mv_-$) then occur in the same direction as the applied field $B$, resulting in a net monopole current $J = J_+ + J_-$. A central new ingredient is that the symmetry derived constraints on spin flips included in the bSM can block monopole motion at local termini, as represented by the monopoles without arrow in Fig. 1A. A monopole is considered to be at a terminus if its preferred motion under the influence of the applied field is blocked (SI *Appendix* section II). The consequences on, and signatures for, non-equilibrium dynamics and such monopole transport theory are entirely unexplored.

To address these issues, here we consider bSM monopole current dynamics in the two essential cases: (a) currents driven by an *instantaneous change* in the applied magnetic field, and (b) currents driven by a *sinusoidal* magnetic field modulation. With a combination of bSM Monte Carlo simulations (*SI Appendix*, section I) and analytical effective modeling (*SI Appendix*, section III) we predict stark dichotomous signatures of bSM monopole current dynamics in these regimes, as illustrated in Fig. 1B and 1C. The pivotal microscopic difference between simple SM and bSM dynamics is the presence of termini in the allowed monopole paths in the latter (Fig. 1A), with specific consequences for the monopole current dynamics. In the first case, a step in applied field modifies monopole motion by inducing dichotomous monopole currents that we distinguish as both *reconfiguration* and *polarization* currents, respectively. The first involves rapid short-range reconfigurations of the monopoles and reflects both microscopic energetic and dynamical constraints on monopole motion; the second changes the polarization of the system via diffusive monopole motion over larger distances, and it decays on a characteristic timescale $\tau_\text{P}$. The reconfiguration current is not present with SM dynamics, and decays on a timescale $\tau_\text{R}$ (approximately the microscopic fast spin-flip rate $\tau_\text{fast}$). It can be intuitively pictured as the driven diffusion of monopoles in and out of termini in response to the field. These mechanisms in turn produce a dichotomy of monopole current responses to a sudden change in applied magnetic field: an initial one being a combination of the polarization and the reconfiguration currents; and a slower second one related only to the polarization of the system.

An intuitive picture for the dichotomous currents can already be gleaned from a very simple one-dimensional ladder: imagine a one-dimensional chain along the $x$ direction, each site of which is also connected to two otherwise isolated `termini' sites located diagonally above and below in opposite directions, i.e., in the $\pm(x + y)$ directions (see Fig. S3). Consider mobile monopoles on the ladder, subject to a field of magnitude $B$ pointing along the chain, and undergoing incoherent Monte Carlo type dynamics. For a field in the $+x$ direction, negative (positive) monopoles will thus be trapped in the sites above (below) the chain; and, crucially, vice versa for a field in the $-x$ direction. Trapped monopoles must then overcome an energy barrier to escape their termini. Reversing the field will free trapped monopoles on one side of the ladder and instead drive them to the other side of the chain. The concomitant motion of these particles yields the reconfiguration current $J_{\text{rec}}$, which coexists with a steady state current $J_{\text{ss}}$. Explicitly, at small fields $J = J_{\text{ss}} + J_{\text{rec}} = \frac{B}{3\tau_{\text{fast}}T} + \frac{2B}{3\tau_{\text{fast}}T} e^{-t/\tau_{\text{fast}}}$ (*SI Appendix*, section III). The long decay constant of $J_{\text{ss}}$ in spin ice becomes infinite in absence of a magnetization buildup in the ladder model, while $J_{\text{rec}}$ decays with the short time constant which is the charge hopping time, equivalent to the spin flip time constant $\tau_{\text{fast}}$ in spin ice. Removing the termini removes the reconfiguration current altogether as occurs equivalently with the dichotomous response in simulations using bSM, but not SM, dynamics. Thus, for example, Fig. 1B shows bSM Monte Carlo simulation results for time dependent magnetization $M(t)$ upon step-wise application of magnetic field of strength 30 mT at time $t = 0$. $M_{\text{sat}}$ is the $t \to \infty$ equilibrium value of the magnetization in the presence of the field. The main panel shows results for bSM dynamics at temperatures 1.7, 2.0, 2.4, 3.0, and 4.0 K. Here the dashed grey lines are exponential fits (highlighting the longer polarization time scale), while the inset contrasts the behaviour between bSM and SM dynamics.

In this theory, the response of magnetic monopoles when subject to an oscillating magnetic field should also reflect the dichotomy of polarization and reconfiguration currents. In the limit of small ($f < \tau_R^{-1}, \tau_P^{-1}$) and large ($f > \tau_R^{-1}, \tau_P^{-1}$) driving frequencies, both currents are either in phase or fully out of phase with the driving field. Crucially, however, there is an intermediate range of driving frequencies $\tau_P^{-1} < f < \tau_R^{-1}$, where the polarization and reconfiguration currents are, respectively, out of and in phase with the driving field. This leads to a pronounced feature in the loss angle $\theta(f) = \arctan(\mathrm{Im}J_f/\mathrm{Re}J_f)$ for AC monopole currents $J(f) = \mathrm{Re}J_f + i\mathrm{Im}J_f$ at frequency $f$. This prediction is visible in Fig. 1C, where we show the $\theta(f)$ extracted from Monte Carlo simulations of spin-ice magnetization, $M(t) = M_0 \sin[2\pi ft + \theta(f)]$, in response to an oscillating magnetic field, $B(t) = B_0 \cos(2\pi ft)$, of amplitude $B_0 = 30$ mT applied along the crystal [111] direction. The main figure shows $\theta(f)$ for bSM dynamics at temperatures 0.8, 1.0, 1.3, 1.7, 2.2, 3.0, and 4.0 K. The inset again contrasts the behaviour between bSM and SM dynamics. In SI Section III, we show how this result can also be found on the three-legged ladder model introduced above. These new perspectives and predictions for bSM monopole current dynamics now motivate our experimental studies to detect and quantify any of the dichotomous transport, dissipation and fluctuation phenomena anticipated.

**Simultaneous monopole current control and spectrometer**

The search for such phenomena requires magnetic field driven monopole currents passing through the pickup coil of a SQUID as shown schematically in Fig. 1A. In this situation $\dot{N}_+$ is the rate of positively charged monopoles (red Fig. 1A) driven along the B-field through the loop to the right, and $\dot{N}_-$ the rate of negatively charged monopoles (blue Fig. 1A) driven oppositely. Thus, the net monopole current $J(t)$ through the persistent superconducting ring and the associated rate of change of flux $\dot{\Phi}(t)$ threading that ring are:

$$J(t) = m\dot{N}_+ - (-m)\dot{N}_- = \dot{\Phi}(t)/\mu_0 \qquad (1)$$

because $\pm m \equiv \pm \Phi_\mathrm{m}/\mu_0$ with $\Phi_\mathrm{m}$ the total magnetic flux of each monopole. To maintain the strict flux quantization required by any persistent superconductive circuit, a spontaneous *electrical* supercurrent then appears flowing around the ring opposing the flux generated by the monopole current, and it is this supercurrent that is linked inductively to the input coil of a SQUID (lower panel of Fig. 1A). To achieve such simultaneous monopole current control and high-precision measurement is technically quite challenging, both because the monopole drive field $B(t)$ applies a giant unwanted flux to the sensing coil and because the whole assembly including the superconductive solenoid generating $B(t)$ must achieve a measurement flux-noise level $\delta \Phi \lesssim 10^{-5}\, \varphi_0/\sqrt{\mathrm{Hz}}$ or equivalent field-noise level $\delta B \lesssim 10^{-14}\, \mathrm{T}/\sqrt{\mathrm{Hz}}$. Figure 1A shows a schematic of the system we have developed to meet these specifications (*SI Appendix*, Fig. S10). A continuous superconductive circuit consisting of a pair of opposite chirality pickup coils ($L_\mathrm{p}$ in Fig. 1A) is assembled along the axis of the drive solenoid (white in Fig. 1A) symmetrically about its center point, and connected to the input coil of the SQUID ($L_\mathrm{i}$ in Fig. 1A). Under these circumstances the applied magnetic field $B(t)$ threads virtually no net flux to the SQUID (*SI Appendix*, Fig. S10). However, if a Dy$_2$Ti$_2$O$_7$ sample is introduced to one of the opposite chirality coils (yellow Fig. 1A), any monopole currents $J(t)$ driven by $B(t)$ along the axis of that crystal generates a changing flux $\dot\Phi(t)$ from Eqn. 1, that can be measured with microsecond precision at the SQUID. The output voltage of the SQUID $V_\mathrm{S}(t)$ is related to the flux $\Phi_\mathrm{p}(t)$ threading the pickup coil as $\Phi_\mathrm{S}(t) = (\mathcal{M}_\mathrm{i}/(2L_\mathrm{p} + L_\mathrm{i}))\Phi_\mathrm{p}(t)$; $L_\mathrm{p}$ is the sample pickup coil inductance, $L_\mathrm{i}$ is the total SQUID-input coil inductance, and $\mathcal{M}_\mathrm{i}$ is a mutual inductance to SQUID and

$$V_\mathrm{S}(t) = \gamma \Phi_\mathrm{S}(t) \qquad (2)$$

where $\gamma$ is the total gain of the electronics (*SI Appendix*, section IV). Hence, both DC and AC monopole currents $J(t)$ can be generated along the Dy$_2$Ti$_2$O$_7$ crystal axis by application of $B(t)$ (green Fig. 2A), measured simultaneously with microsecond precision using $\mu_0 \Phi_\mathrm{S}(t) \propto \int J(t)\, dt$ (blue Fig. 2A), as well as the instantaneous power spectral density of magnetic noise $S_M(\omega, T) \propto S_{\Phi_S}(\omega, T) = \gamma^2 S_{V_S}(\omega, T)$, where $S_{V_S}(\omega, T)$ is the output voltage noise spectrum of the SQUID. Our objective is then to use this new spectrometer to search for the dichotomous monopole transport, dissipation and fluctuation phenomena.

**Monopole current timescale dichotomy**

Dy$_2$Ti$_2$O$_7$ crystals, prepared and evaluated for this purpose are inserted into one of the counter wound pickup coils with it a long axis along the crystal [351] direction (yellow Fig. 1A & *SI Appendix*, section V) and cooled on board a custom built ultra-low vibration refrigerator mounted on the isolated floor-slab of an ultra-low vibration laboratory (*SI Appendix*, section IV). Figure 2A shows a typical example of a monopole current control generation and detection experiment. The green trace shows the magnetic field $B(t)$ as a function of time while the blue curve is the time dependence of flux $\Phi_S(t)$ at the SQUID. Axially, there are two possible field $B_\pm(t)$ and current $\dot\Phi_{S\pm}(t)$ directions here, and both are studied throughout. A typical example of measured $\log \Phi_S(t)$ evolution beginning 200 μs after monopole current initiation (MCI) is shown in Fig. 2B. These unprocessed data reveal two distinct monopole current regimes. At long times $t > 600$ μs we observe a well-defined time constant $\tau_2$ for monopole current flow as indicated by the dashed straight line fit to $\log \Phi_S(t)$. At short times $t < 600$ μs from MCI, a transition occurs to a shorter time constant $\tau_1$ for monopole current flow. To analyze all such curves at all temperatures we fit $\log \Phi_S(t, T) = C(T) - t/\tau_2(T)$ for times $600$ μs $< t < 1200$ μs and derive $\tau_2(T)$ and $C(T)$ for all examples with fit quality factor $R^2 > 0.99$ (*SI Appendix*, section VI & Fig. S12). The slow-decaying monopole current $\Phi_2(t, T)$ data represented by these fits are subtracted in the time range $200$ μs $< t < 1200$ μs from the measured $\Phi_S(t, T)$ to yield the residual fast-decaying monopole current $\Phi_1(t, T)$ data, such that $\Phi_S = \Phi_1 + \Phi_2$. Figure 2D upper panel shows the resulting $\Phi_1(t, T)$ for fast-decaying currents while the simultaneous $\Phi_2(t, T)$ for slow-decaying currents are shown in the lower panel, both for positive *B*-field direction. Figure 2E shows the equivalent analyzed monopole current data for negative *B*-field direction. In both cases, the fast-decaying monopole currents have ceased for $t > 600$ μs from MCI, while the slow-decaying monopole currents persist for many milliseconds.

**Monopole dissipative loss angle and noise dichotomy**

To probe the possible presence of this dichotomy also in the alternating magnetic monopole current (AC) transport characteristics, we study sinusoidal monopole current generation and detection, a typical example of which is shown in Fig. 3A. The green trace shows the applied magnetic field $B_0(t) = B\cos(2\pi f t)$ as a function of time while the blue curve is the simultaneously measured time dependence of flux $\Phi_S(t)$ at the SQUID. AC monopole currents are studied for $10 \leq f \leq 5000$ Hz (*SI Appendix*, section VII). The monopole current $J(t) = \text{Re}J_f\cos(2\pi f t) + i\text{Im}J_f\sin(2\pi f t)$ is determined by using a lock-in amplifier to measure $\Phi_S(t) = \text{Re}\Phi_S\cos(2\pi f t) + i\text{Im}\Phi_S\sin(2\pi f t)$ for all $f$. The measured flux is linked to the monopole current as $\text{Re}J_f \propto -2\pi f \text{Im}\Phi_f$ and $\text{Im}J_f \propto 2\pi f \text{Re}\Phi_f$. The measured $\text{Re}J_f$ and $\text{Im}J_f$ for all measured temperatures are shown in Fig 3B. As with any harmonic, charged, non-momentum-conserving fluid dynamics, the dissipation characteristics are established using the loss angle $\theta_d(f) \equiv \arctan(\text{Im}J_f/\text{Re}J_f)$. In Fig 3C we show measured $\theta_d(f)$ for AC monopole currents in Dy$_2$Ti$_2$O$_7$ from all temperatures studied, while the inset shows the frequency $f_d$ associated with a change in the frequency dependence of $\theta_d(f)$. By estimating the frequency $f_d$ which leads to the local maximum in $d\theta_d/df$, AC time constant $\tau_d$ is illustrated in Fig. 4A, showing a consistent transition at frequency $f_d \cong 1.8$ kHz for all temperatures studied.

Finally, we search for a change in the magnetic noise intensity that one might anticipate to occur when the fast-decaying reconfiguration currents have ceased[23]. The flux noise intensity $\sigma_1^2(t) = \langle\Phi_S(t)^2\rangle$ is measured as soon after MCI as possible thus during the epoch of fast-decaying currents $0 < t < 600$ μs. Subsequently, $\sigma_2^2(t) = \langle\Phi_S(t)^2\rangle$ is measured throughout the epoch of slow-decaying polarization currents $600$ μs $< t <$ $2000$ μs. Their ratio $\sigma_1^2/\sigma_2^2$ is shown in Fig. 4B exhibiting a constant value near 1.5 for all temperatures (*SI Appendix*, section VIII). The increase of the monopole noise in the reconfiguration current regime is also observed in time-dependence of variance (*SI Appendix*, Figure S13). Hence the magnetic noise power does indeed drop steeply when the fast-decaying currents disappear at $t \approx 600$ μs.

**Discussion**

The preeminent feature of these experimental findings is the presence of two distinct timescales (Fig. 2C & Fig. 4C). This is naturally accounted for by the different decay times within the bSM monopole dynamics theory of polarization and reconfiguration currents as in Fig. 1B, C (*SI Appendix*, section I & IX), implying the time scales $\tau_1$ and $\tau_2$ measured in experiments relate directly to the theoretically predicted $\tau_R$ and $\tau_P$, respectively. Indeed, this agreement between the bSM monopole current theory and experiments exists in considerable detail: the ratio of the two timescales in both Monte Carlo simulation and experiment hovers around a value of $\frac{\tau_1}{\tau_2} \approx \frac{1}{3}$ in the high temperature window above 2 K (*SI Appendix*, section X). The values differ by approximately 1/3 in the high temperature part of our measurement regime, but that $\tau_2$ grows significantly faster upon lowering of the temperature in MC simulations than observed in experiments. Neither the monopole density at high temperature (*SI Appendix*, section XI) nor the demagnetization factor from the sample geometry (*SI Appendix*, section XII) could give rise to this observation. The origin of this theoretical discrepancy remains to be understood and is further discussed in *SI Appendix*, section XIII. A characteristic feature in the experimentally measured loss-angle $\theta_\text{d}(f,T)$ appears at frequencies $\tau_2^{-1} \lesssim f < \tau_1^{-1}$, as predicted theoretically (Fig. 3C), although the phenomenon is discernible at higher temperature than observable by bSM Monte Carlo simulations in Fig. 1C. Finally, the magnetic noise power drops steeply after time $t \approx 600$ μs (Fig. 4B), indicating suppression of noise once the reconfiguration currents have decayed away on the short timescale $\tau_1$. Thus, the dichotomous monopole transport theory is highly consistent with the observed monopole current phenomenology. Moreover, if there are multiple microscopic spin-flip time constants[23] potentially subtending all dynamical processes in spin-ice[24,25], the dichotomous monopole current theory presented here is quantitatively consistent with the phenomenology of $Dy_2Ti_2O_7$ for a fast time constant of approximately 90 μs.

To summarize: the recent discovery of dynamical fractal trajectories underpinning equilibrium monopole motion in real materials[23] motivated development of our new theory for magnetic monopole currents in spin-ice. Using a combination of analytical effective theory and bSM Monte Carlo simulations, this theory predicts a dichotomy of the driven monopole current dynamics with characteristic signatures in measurable quantities. State-of-the-art monopole current spectroscopy developed to explore these predictions discovers strong dichotomy in the monopole current response of $Dy_2Ti_2O_7$ to sudden changes in applied magnetic fields, and in the dissipative loss-angle in response to AC fields, as well as in the magnetization noise. The consequence is a novel and accurately predictive, atomic-scale mechanism for magnetic monopole currents in spin-ice. Uniquely, as in the case of equilibrium fractal dynamics, this involves the geometry of allowed trajectories subject to energetic and dynamical constraints, in particular the termini to monopole motion. This new paradigm is of self-generated dynamical heterogeneities in quantum magnetic transport which are not caused solely by energetic constraints (e.g., interactions), nor by geometric constraints (e.g., excluded volume effects), nor by quenched disorder. Instead, we demonstrate how the interplay of atomic-scale spin configurations maintaining lattice symmetry nevertheless lead to a distribution of dynamical time scales within an otherwise perfectly crystalline material. Ultimately, this mechanism reveals a new avenue for discovery and exploration of self-constrained dynamics and quantum transport mechanisms in other classes of quantum magnets.

# FIGURES

**Figure 1 Simultaneous Monopole Current Control and Measurement Spectrometer**

A. Conceptual representation of magnetic field driven monopole current $J(t)$ passing through a superconducting loop (yellow). Positive charged monopoles (red) are driven to the right by an applied field $B$, and negatively charged (blue) to the left. These rapid monopole currents are occasionally terminated when the spin-flip rate is suppressed by specific local spin conformations, magnified within the smaller panels at left and right. Lower panel: Conceptual design of simultaneous monopole current control and measurement system based on direct high-precision SQUID sensing of the monopole current $J(t) = (d\Phi/dt)/\mu_0$

B. Monte Carlo simulation results of the magnetic response when an external magnetic field of strength 30 mT is suddenly applied at time $t = 0$. $M_{sat}$ is the equilibrium value of the magnetization in the presence of the field. The main panel shows results for bSM dynamics at temperatures 1.7 K, 2.0 K, 2.4 K, 3.0 K, and 4.0 K (from upper to lower lines). The dashed grey lines are exponential fits (highlighting the longer polarization time scale). The inset contrasts the behavior at shorter times and temperature 1.7 K between bSM (dark blue) and SM (orange) dynamics, and the corresponding dashed line fit (highlighting the shorter polarization time scale in SM dynamics). The fast and slow contributions are plotted separately in Fig. S19.

C. The loss angle $\theta(f)$ extracted from Monte Carlo simulations of spin-ice magnetization, $m(t) = m_0 \sin[2\pi f t + \theta(f)]$, in response to an oscillating magnetic field, $B(t) = B_0 \cos(2\pi f t)$, of amplitude $B_0 = 30$ mT and frequency $f$ applied along the [111] direction. The main figure shows the loss angle for bSM dynamics at temperatures 0.8 (dark blue), 1.0, 1.3, 1.7, 2.2, 3.0, and 4.0 (dark red) K. A bump-like feature is clearly evident at low temperatures, induced by a reconfiguration current contribution originating from the presence of termini in the emergent dynamical fractal, as explained in the main text. The inset shows the loss angle at 0.8 K (dark blue) and 1.7

K (yellow) for both bSM (circles and solid lines) and SM (squares and dotted lines) dynamics; notice the absence of the characteristic loss-angle feature in the latter.

**Figure 2 Magnetic Monopole Current Dichotomy in Dy₂Ti₂O₇**

A. Typical example of monopole current control generation and detection system in operation. The green trace shows the magnetic field as a function of time while the blue curve is the time dependence of flux $\Phi_S(t)$ measured at the SQUID. Monopole current initiation (MCI) time marked by the blue sign is set to 0 for each current transient.

B. Typical example of $\log \Phi_S(t)$ evolution beginning 200 $\mu s$ after MCI at $t = 0$. Since monopole current is $J(t) = (d\Phi/dt)/\mu_0$ and $\Phi(t) \propto \Phi_S(t)$, these unprocessed data reveal two distinct monopole current regimes. At long times there is a well-defined time constant $\tau_2$ for monopole current flow as indicated by the dashed straight line fit. At short times $t < 600\ \mu s$ from MCI, a transition occurs to a much shorter time constant $\tau_1$ for monopole current flow.

C. Measured $\log \Phi_S(t)$ evolution beginning 200 $\mu s$ after MCI for all temperatures studied and for both positive and negative magnetic field directions. For all transients at long times the time constant $\tau_2(T)$ is measured by a straight line fit. At short times $t < 600\ \mu s$ after MCI for all these transients, a transition occurs to a faster time constant $\tau_1(T)$ for monopole current decay.

D. Extracted $\Phi_1(t)$ for fast-decaying currents and positive *B*-field direction. These fast-decaying currents have ceased for $t > 600\ \mu s$ from MCI. Simultaneous data in the lower panel shows extracted $\Phi_2(t)$ for slow-decaying currents and positive *B*-field direction.

E. Extracted $\Phi_1(t)$ for fast-decaying currents and negative *B*-field direction. These fast-decaying currents have ceased for $t > 600\ \mu s$ from MCI. Simultaneous data in the lower panel shows extracted $\Phi_2(t)$ for slow-decaying currents and negative $B$-field direction. The phenomenology is indistinguishable from that in **d**.

### Figure 3 AC Magnetic Monopole Loss Angle in Dy$_2$Ti$_2$O$_7$

A. Typical example of sinusoidal monopole current generation and detection. Green trace shows the magnetic field $B_0(t) = Bcos(2\pi ft)$ as a function of time while the dark blue curve is the measured time dependence of flux $\Phi_S(t)$ at the SQUID and the light blue curve is $d\Phi_S(t)/dt$. This field modulation experiment is carried out for $10\ Hz \leq f \leq 5000\ Hz$.

B. From A and with $J(t) \propto 1/\mu_0(d\Phi_S(t)/dt)$, monopole current $J(t) = ReJ_f cos(2\pi ft) + iImJ_f sin(2\pi ft)$ is determined by using a lock-in amplifier to measure $\Phi_S(t) = Re\Phi_f cos(2\pi ft) + iIm\Phi_f sin(2\pi ft)$ for all $f$. The consequent in-phase current $ReJ_f \equiv -1/\mu_0(2\pi f Im\Phi_f)$ and out-of-phase current $ImJ_f \equiv 1/\mu_0(2\pi f Re\Phi_f)$ are show for all temperatures measured. The corresponding rate $\dot{N}$ of the number of monopoles passing through the superconducting loop is estimated by $\dot{N} = \dot{\Phi}_S(t)/\mu_0 m$. The practical rate $\dot{\Phi}_p$ can be converted through the coefficient $M_i/(2L_p + L_i)$.

C. Measured $\theta_d(f) = arctan(ImJ_f/ReJ_f)$ from all temperatures studied. Inset shows that a dissipation transition occurs at $f_d \cong 1.8\ kHz$.

### Figure 4 Dichotomous Monopole Currents, Dissipation and Noise

A. The transition point $\tau_d = 1/2\pi f_d$ where the dissipative process alteration occurs at $f_d \approx 1.8\ kHz$ from all temperatures studied.

B. Ratio of monopole current driven magnetization noise intensity $\sigma_1^2$ for fast-decaying currents $t < 600\ \mu s$ from MCI to magnetization noise intensity $\sigma_2^2$ for slow-decaying currents during $600\ \mu s < t < 1200\ \mu s$. This ratio is constant near 1.5 for all temperatures.

C. Experimentally determined monopole current relaxation time constants from all temperatures studied and both field application directions. Measured slow-decaying monopole current time constant $\tau_2$ in solid red; fast-decaying monopole current time constant $\tau_1$ in solid blue.

**Acknowledgements**: We acknowledge and thank E.-A. Kim, S.A. Kivelson, A.P. Mackenzie, and S. Sondhi for key discussions and guidance. J.C.S.D. acknowledges support from the Moore Foundation's EPiQS Initiative through Grant GBMF9457. C.-C.H. and J.C.S.D. acknowledge support from the European Research Council (ERC) under Award DLV-788932. H.T. and J.C.S.D. acknowledge support from the Royal Society under Award R64897. JCSD and FJ thank the MPI-CPFS for support. J.D., C.Carroll, J.W., C.D. and J.C.S.D. acknowledge support from Science Foundation of Ireland under Award SFI 17/RP/5445. C.Carroll acknowledges support from Irish Research Council under Award GOIPG/2023/4014. G.M.L. acknowledges support from the Natural Sciences and Engineering Research Council (Canada). S.J.B. acknowledges support from UK Research and Innovation (UKRI) under the UK government's Horizon Europe funding guarantee (Grant No. EP/X025861/1). C.Castelnovo acknowledges support in part from the Engineering and Physical Sciences Research Council (EPSRC) grants No. EP/P034616/1, EP/V062654/1 and EP/T028580/1. J.N.H and R.M. acknowledge support in part from the Deutsche Forschungsgemeinschaft under grants SFB 1143 (project-id 247310070), and the cluster of excellence ct.qmat (EXC 2147, project-id 390858490).

**Author Contributions:** J.C.S.D., R.D., C.C. and R.M. conceived the project. G.L. synthesized and characterized the samples; J.D., J.W., C. Carroll, H.T., R.D. and C.-C.H. developed spin spectrometry techniques; C. C., J.N.H. and R.M. developed theory and interpretation; S.J.B. provided theoretical guidance; C.-C.H., F.J. and H.T. carried out experimental measurements; C.D., F.J., H.T. and C.-C.H. developed and carried out the data analysis. J.C.S.D., S.J.B. , C.C. and R.M. supervised the research and wrote the paper with key contributions from J.N.H., F.J., H.T. and C.-C.H. The manuscript reflects the contributions and ideas of all authors.

**Author Information** Correspondence and requests for materials should be addressed to jcseamusdavis@gmail.com

# Figure 1

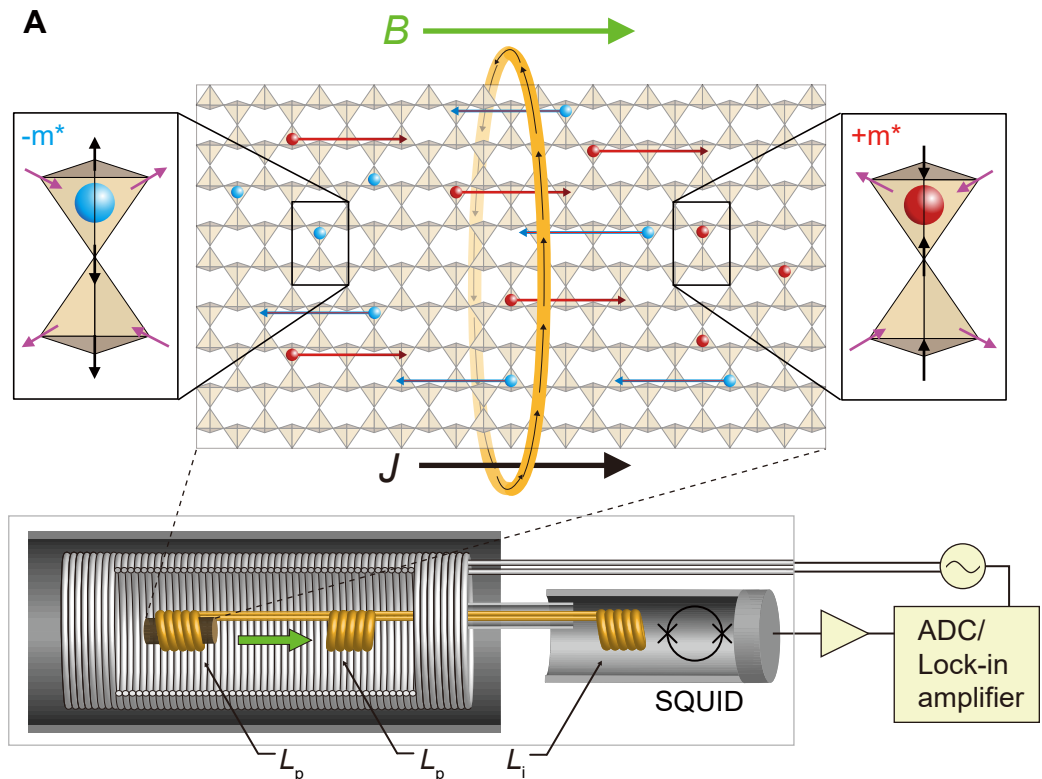
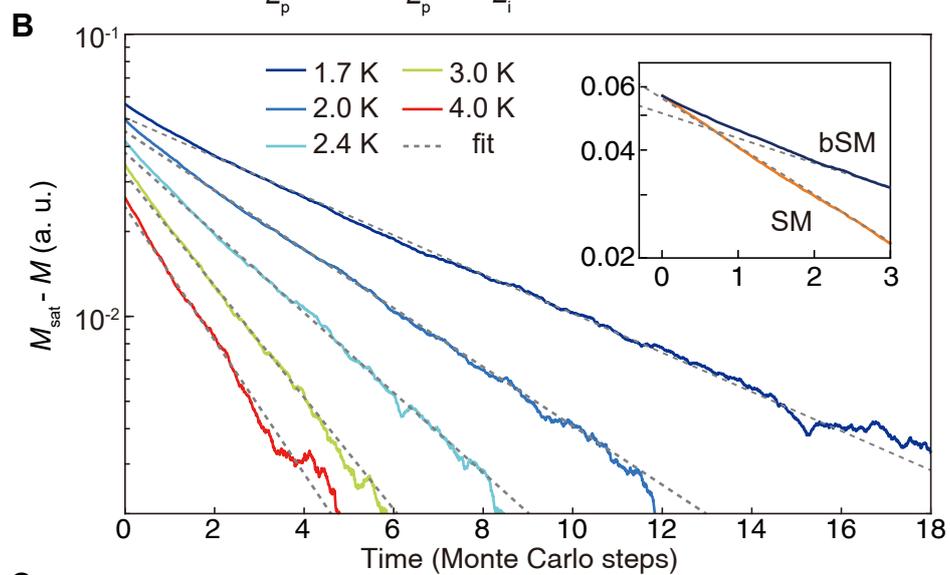
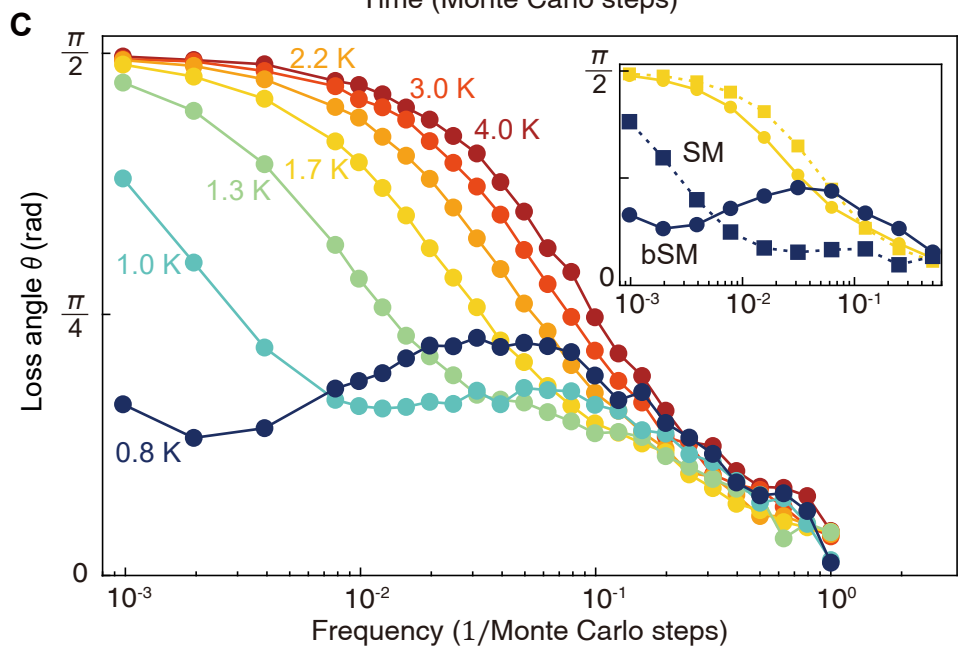

Figure 2

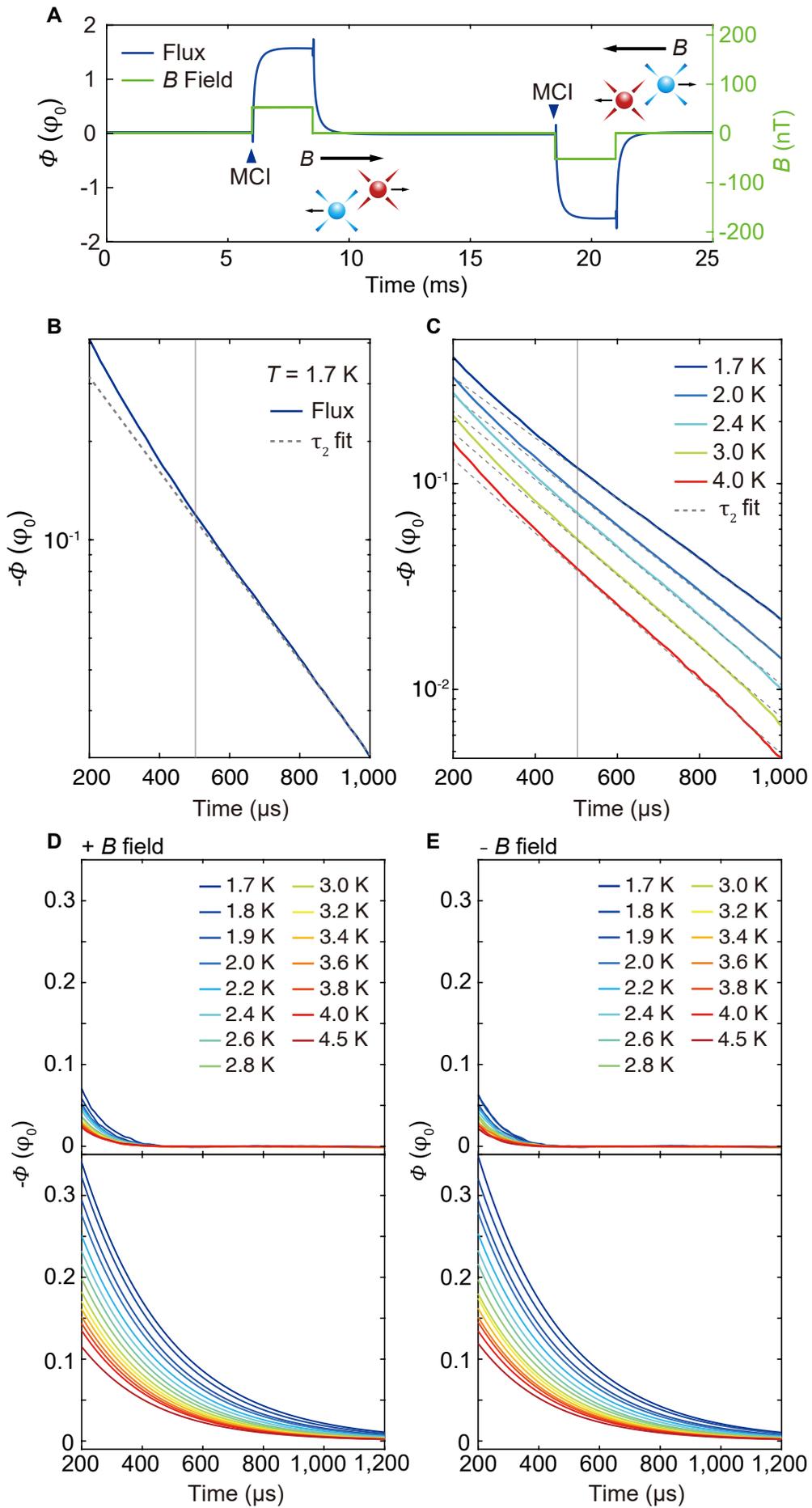

# Figure 3

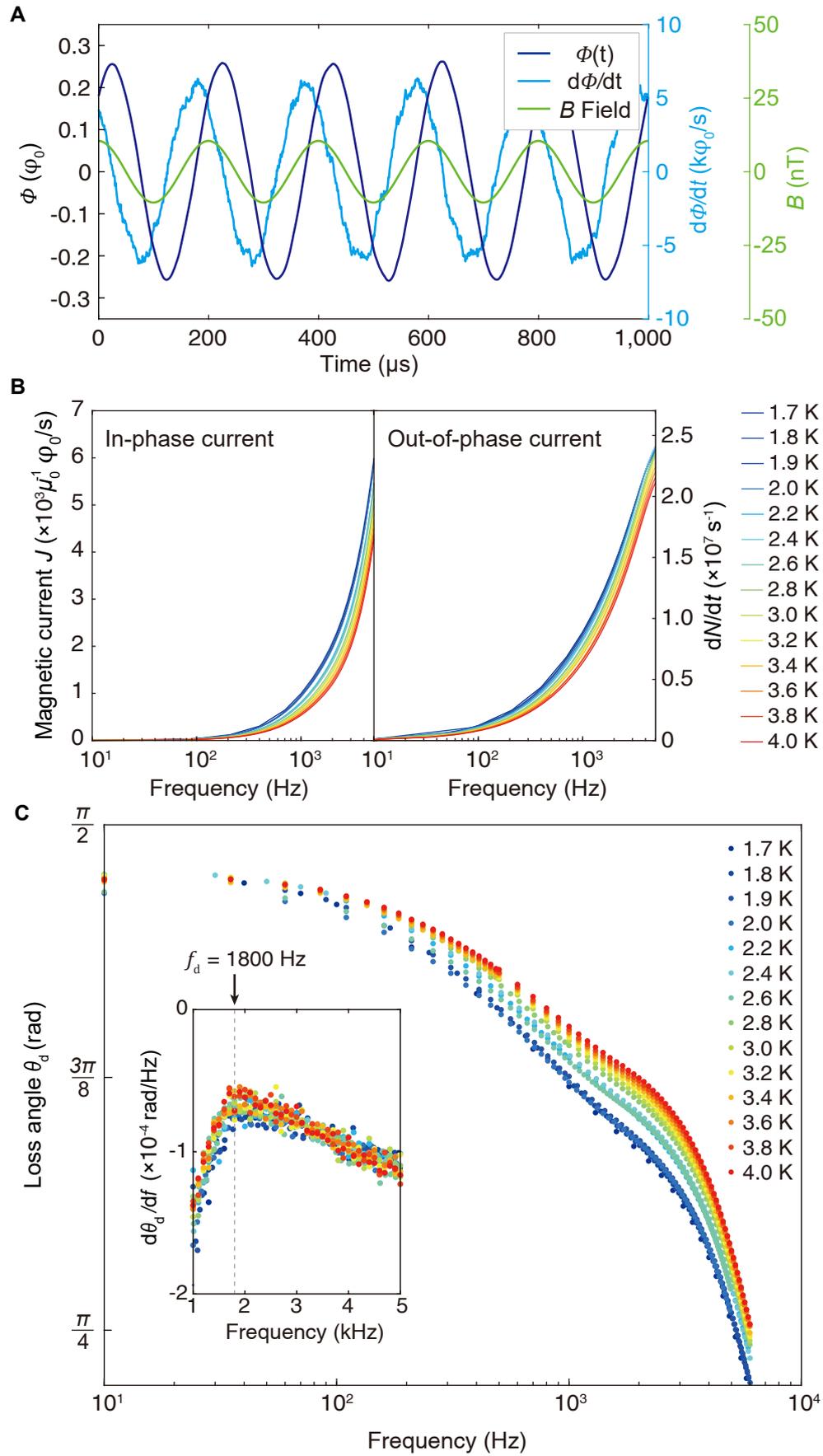

Figure 4

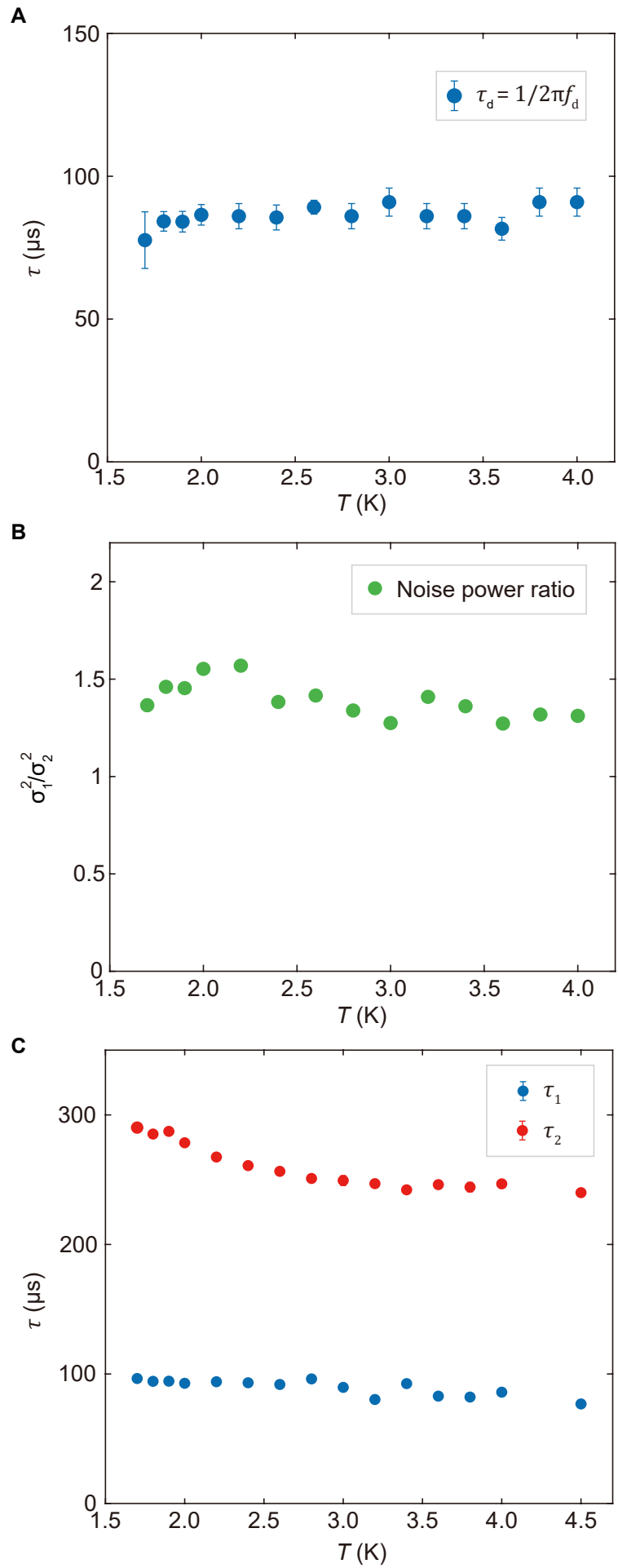

**Supplementary Information for**

**Dichotomous Dynamics of Magnetic Monopole Fluids**

Chun-Chih Hsu, Hiroto Takahashi, Fabian Jerzembeck, Jahnatta Dasini, Chaia Carroll,
Ritika Dusad, Jonathan Ward, Catherine Dawson, Sudarshan Sharma, Graeme Luke,
Stephen J. Blundell, Claudio Castelnovo, Jonathan N. Hallén, Roderich Moessner
and J.C. Séamus Davis

## (I) Numerical method

All numerical results presented in this paper were obtained using classical Monte Carlo simulations with the Metropolis algorithm, of pyrochlore spin ice systems of $L \times L \times L$ cubic unit cells containing 16 spins each. Periodic boundary conditions were used throughout, and the Ewald summation technique was used to include long-ranged dipolar interactions (26). The results presented in the main text were obtained using a Hamiltonian consisting of long-range dipolar interactions and short-range exchange between first, second, and third-neighbour spins:

$$\mathcal{H}_{\text{OP}} = Da^3 \sum_{i<j} \left[ \frac{\vec{S}_i \cdot \vec{S}_j}{r_{ij}^3} - \frac{3(\vec{S}_i \cdot \vec{r}_{ij})(\vec{S}_j \cdot \vec{r}_{ij})}{r_{ij}^5} \right]$$
$$+ J_1 \sum_{\langle i,j \rangle} \vec{S}_i \cdot \vec{S}_j + J_2 \sum_{\langle i,j \rangle_2} \vec{S}_i \cdot \vec{S}_j$$
$$+ J_3 \sum_{\langle i,j \rangle_3} \vec{S}_i \cdot \vec{S}_j + J_{3'} \sum_{\langle i,j \rangle_{3'}} \vec{S}_i \cdot \vec{S}_j \quad (S1)$$

The dipolar interaction strength is $D = 1.3224$ K/$a^3$, and the exchange strengths are $J_1 = 3.41$ K, $J_2 = 0.0$ K, $J_3 = -0.00466$ K, and $J_3' = 0.0439$ K. This Hamiltonian was obtained as a combined fit to neutron scattering, magnetic susceptibility, and specific heat measurements on Dy$_2$Ti$_2$O$_7$ in Ref. (27). As is conventional for spin ice, the spin variables $\vec{S}_i$ are unit vectors constrained to point along the local pyrochlore easy-axes. In simulations with $\mathcal{H}_{\text{OP}}$ – all simulations presented in the main text – we used $L = 10$, i.e., a cubic system consisting of 16,000 spins.

The dynamics were simulated using single spin flip updates. Each update consists of selecting a spin at random and computing the energy cost $\Delta E$ associated with flipping the spin. $\Delta E$ contains both the cost due to spin interactions – encoded in the Hamiltonian – as well as the energy cost due to the applied field, which is $\mu \vec{S}_i \cdot \vec{B}$ for spin $i$, where $\mu \approx 10\mu_B$ is the magnetic moment of the dysprosium ions. The update is then accepted with probability $P$ = min [1, exp(-$\Delta E/T$)] in all cases, if SM dynamics are used. If bSM dynamics are used, the transverse magnetic field from the six nearest-neighbour spins is also considered, and the proposed move is rejected when this transverse field is vanishingly small (as explained in Ref. (*23,28*). A Monte Carlo step consists of *N* attempted spin flips, where *N* is the number of spins in the system. One such step corresponds to the time $\tau_\text{fast}$, where $1/\tau_\text{fast}$ is a temperature independent rate of attempted spin flips in the dynamical model.

In all simulations the system is first cooled down in zero field to the measurement temperature using simulated annealing. A field is then applied along a given direction and the magnetization is measured (along the same direction) In the case of a suddenly applied field, shown in Fig. 1B, we average the magnetization over 150 independent histories for different random number seeds. In the case of an oscillating applied field ($B$(t) = B$_0$cos($2\pi ft$)), the system is first allowed to reach its steady state (usually about 50-100 periods suffice), and then the loss angle *θ(f)* at frequency *f* is estimated by fitting a function $A$sin($2\pi ft+\theta$), with fitting parameters A and *θ*, to the magnetization measured over 10 periods and averaged over 20 independent simulations with different random number seeds.

The magnetic field strengths used in our numerics are necessarily larger than those used in the experiments, because of the larger statistical noise incurred in the necessarily smaller system sizes accessible in simulations. For example, at 2 K and applied 100 nT magnetic field, the probability of flipping a spin to align with the field is approximately $10^{-7}$ times more likely than flipping it to anti-align with the field. Such differences are readily detectable in experimental samples with $\sim 10^{23}$ spins and typical SQUID measurement accuracy, but cannot be realistically observed in simulations where we only have access to $\sim 10^4$ spins. We therefore had to consider larger field strengths, and in this work, we used fields of the order of 10 mT. We verified that this is a weak enough field for the system to remain in the linear response regime (see Sec. VIII), and therefore we expect our results to be representative of the behaviour at fields of 100 nT, if we had the computational power to simulate them. We have also verified that the simulation results show only very weak dependence on field application direction, and all results presented here hold for an arbitrary field direction.

## (II) Monopoles in termini

In simulations one can directly measure the proportion of monopoles that are in termini at any given time, with respect to a given field driving direction. This information helps form a more insightful picture of how the reconfiguration currents arise and behave.

We define a monopole as being in a positive terminus if it cannot move in its preferred direction for an applied field in the [100] direction and in a negative terminus if it cannot move in its preferred direction for an applied field in the [-100] direction. We call the proportion of all monopoles that are in positive or negative termini $\gamma_+$ and $\gamma_-$. Note that this definition counts monopoles constrained to a single site (a few percent of all monopoles) as being in simultaneous termini in both directions.

A sudden applied magnetic field along [100] initially increases $\gamma_+$ and decreases $\gamma_-$ by driving monopoles in and out of the respective termini (see Fig. S1). This is a direct manifestation of the reconfiguration current. After a time $\sim \tau_R$, $\gamma_+$ and $\gamma_-$ hit their respective maximum and minimum. Configurations with monopoles in negative termini are energetically favoured and configurations with monopoles in positive termini are energetically penalised in the field. After the initial fast response due to reconfiguration ($\tau_R$), there follows a slow decrease of $\gamma_+$ and a corresponding slow increase of $\gamma_-$ towards a final state with $\gamma_- > \gamma_+$ dictated by the energetic preference between the two introduced by the applied magnetic field. This second polarisation response occurs on a longer time scale $\tau_P$.

Let us now consider the case of driving with a sinusoidal applied field, in the frequency regime $f > \tau_P^{-1}$ (i.e., the field is changing too quickly for the system to polarise). While the magnetisation is out of equilibrium, we find that $\gamma_+$ and $\gamma_-$ remain approximately in equilibrium up to larger frequencies (up to $f \lesssim \tau_P^{-1}$), and a component of the magnetisation therefore oscillates in phase with the field, driven by the reconfiguration current. On the other hand, the magnitude of the reconfiguration current drops when the frequency is reduced much below $\tau_P^{-1}$, and we observe a significant reconfiguration contribution to the oscillatory response only at intermediate frequencies ($\tau_R^{-1} \gtrsim f > \tau_P^{-1}$). The behaviour of the termini populations $\gamma_-$ and $\gamma_+$ in response to an oscillating field is illustrated in further detail in Fig. S2.

## (III) Effective models: Jagged and uniform three-leg ladder

In this section we present two analytically tractable effective models to demonstrate how charged particles moving on a system with termini behave in an applied field. The setups we consider are infinite three-leg ladders where the sites are connected in two different ways, as illustrated in Fig. S2. In each case, we place a charged random walker on the ladder and study its dynamics in presence of an applied field. We denote the probability that a walker is on the central leg by $p_0$, and that it is on one of the lower/upper legs by $p_-/p_+$. Normalisation gives $p_- + p_0 + p_+ = 1$. Also, let us assume that the probability that a jump is attempted in a small time $\delta t$ is $\delta t/\tau_0$, where $\tau_0$ is some characteristic time scale. Mapping this to spin ice $\tau_0$ is equivalent to $\frac{1}{4}\tau_{\text{fast}}$, as there are four surrounding spins whose flip can move a monopole. For simplicity, we choose to implement the same Metropolis dynamics used in our Monte Carlo spin ice simulations. Under the influence of an applied magnetic field $B$ along the ladder, chosen without loss of generality to be positive in the $+x$ direction (shown in Fig. S3), we stochastically move a charge between two adjacent sites with probability $Q_+ = \min[1, \exp(B/T)]$, if the move is in the $+x$ direction, and with probability $Q_- = \min[1, \exp(-B/T)]$, if the move is in the $-x$ direction.

Note that the jagged ladder has four bonds connected to every site on the central leg, but only one bond connected to every site on the upper and lower legs. This results in termini – sites where the only available move for a walker is against the field direction. The dense ladder has six bonds connected to every site on the central leg, and two bonds connected to every site on the upper and lower legs. It therefore has no termini.

### Jagged ladder

On the central leg of the jagged ladder, a charge can move in four possible directions: "up", "down", "left", or "right", with equal probability; on the upper / lower leg, the only allowed moves are "down"/"up", respectively.

Since $p_0 = 1 - p_+ - p_-$, we can describe the dynamics using only $p_+$ and $p_-$. One can write down the rates of change for these probabilities as

$$\frac{dp_-}{dt} = \frac{1}{4\tau_0}[(1 - p_- - p_+)Q_- - p_- Q_+] \tag{S2}$$

and

$$\frac{dp_+}{dt} = \frac{1}{4\tau_0}[(1 - p_- - p_+)Q_+ - p_+Q_-] . \tag{S3}$$

The equilibrium values, defined by $\frac{dp_\pm^{(eq)}}{dt} = 0$, are given by

$$p_\pm^{(eq)} = \frac{e^{\pm B/T}}{1 + e^{-B/T} + e^{B/T}} , \tag{S4}$$

which satisfy the expectation that $p_+ = p_- = p_0 = 1/3$ if $B = 0$, $p_- \to 1$ if $B \to -\infty$, and $p_+ \to 1$ if $B \to +\infty$. Fig. S4 shows the flow of $p_+$ and $p_-$ for different initial conditions and a constant $B/T$. $p_+$ and $p_-$ flow to the equilibrium values given above for any possible initial configuration.

One can then write the total current $J = \langle dx/dt \rangle$ as

$$J_{\text{jagged}} = \frac{1}{4\tau_0}[(2 - p_- - 2p_+)Q_+ - (2 - 2p_- - p_+)Q_-] . \tag{S5}$$

In our case, it is useful to express it as the sum of a steady state and a transient component. The steady state current results from the field moving the walker along the ladder and is given by

$$J_{\text{ss}} = \frac{p_0}{4\tau_0}\left(1 - e^{-\frac{|B|}{T}}\right)\text{sign}(B) . \tag{S6}$$

We understand this as being analogous to the polarisation current in spin ice. However, by the simplicity of its construction, our ladder does not actually polarise and charges can move along it indefinitely; unlike spin ice, the ladder supports a DC current. To stress this important difference, we shall not refer to $J_{\text{ss}}$ as the polarisation current.

The transient current is given by

$$\begin{aligned} J_{\text{rec}} &= \frac{dp_+}{dt} - \frac{dp_-}{dt} \\ &= \frac{1}{4\tau_0}[(1 - p_+)Q_+ - (1 - p_-)Q_-] . \end{aligned} \tag{S7}$$

It appears when the populations of the three legs are out of equilibrium, and there is net flow between them. We understand this as being equivalent to the reconfiguration of monopoles on their local clusters in spin ice, and we therefore refer to the transient current on the jagged ladder as reconfiguration current. Noting that $Q_-/Q_+ = \exp(-B/T)$, one can show that $J_{\text{rec}} = 0$ for

$$\frac{1-p_+}{1-p_-} = e^{-B/T}. \tag{S8}$$

For a given value of $B/T$ the solutions to $J_{\text{rec}} = 0$ thus form a straight line through the $p_+$-$p_-$ plane, with one specific solution given by $p_+^{(\text{eq})}$ and $p_-^{(\text{eq})}$ (see Eq. (S4)):

$$\frac{1-p_+^{(\text{eq})}}{1-p_-^{(\text{eq})}} = \frac{1+e^{-B/T}}{1+e^{+B/T}} = e^{-B/T}. \tag{S9}$$

$p_+^{(\text{eq})}$ and $p_-^{(\text{eq})}$ is the only point for which $J_{\text{rec}} = 0$ and $\frac{dJ_{\text{rec}}}{dt} = 0$; as $p_+$ and $p_-$ flow to this point the reconfiguration current will thus decay to zero.

**Dense ladder**
On the central leg of the dense ladder, a charge can move in one of six possible directions: "up left", "up right", "down left", "down right", "left", or "right", with equal probability; on the upper/lower leg, the allowed moves are "down left" or "down right"/"up left" or "up right", respectively

The rates of change of the probabilities now take a completely symmetric form

$$\frac{dp_-}{dt} = \frac{1}{6\tau_0}[(1-p_- -p_+)(Q_- + Q_+) - p_-(Q_- + Q_+)] \tag{S10}$$

and

$$\frac{dp_+}{dt} = \frac{1}{6\tau_0}[(1-p_- -p_+)(Q_- + Q_+) - p_+(Q_- + Q_+)]. \tag{S11}$$

In this case the equilibrium values are field independent and given by $p_+ = 1/3$ and $p_- = 1/3$. If we assume that the ladder is in equilibrium when the field is first applied, the populations do not evolve at all in time, and there is no reconfiguration current. The total current is then given by

$$J_{\text{dense}} = \frac{1}{6\tau_0}[(3 - 2p_- - 2p_+)Q_+ - (3 - 2p_- - 2p_+)Q_-]$$

$$= \frac{1}{6\tau_0}(3 - 2p_- - 2p_+)(1 - e^{-|B|/T})\text{sign}(B)$$

$$= \frac{1}{6\tau_0}\frac{5}{3}\left(1 - e^{-\frac{|B|}{T}}\right)\text{sign}(B). \tag{S12}$$

**Sudden applied field**

If we suddenly turn on a weak magnetic field ($B \ll T$) at time $t = 0$, the steady state currents arising to first order in $B/T$ are

$$J_{ss} \approx \frac{B}{12\tau_0 T} \tag{S13}$$

and

$$J_{dense} \approx \frac{5B}{18\tau_0 T}. \tag{S14}$$

The reconfiguration current on the jagged ladder is somewhat more complicated. In the weak-field limit the equilibrium configuration becomes

$$p_\pm^{(eq)} \approx \frac{1 \pm B/T}{3}. \tag{S15}$$

Define for convenience $p_+(t) = 1/3 + \delta_+(t)$ and $p_-(t) = 1/3 + \delta_-(t)$. To first order in $B/T$ one can show that $\delta_+(t) = -\delta_-(t) = \delta(t) \geq 0$ (for $B \geq 0$). From Eq. (S3), or equivalently from Eq. (S2), one then finds

$$\frac{d\delta}{dt} \approx \frac{1}{4\tau_0}\left[\frac{1}{3} - (1/3 + \delta)\left(1 - \frac{B}{T}\right)\right]$$

$$\approx \frac{1}{4\tau_0}\left[\frac{B}{3T} - \delta\right], \tag{S16}$$

where the term proportional to $\delta B/T$ can be neglected as it is of order $(B/T)^2$. Solving Eq. (S16) with the initial condition $\delta(t = 0) = 0$ one arrives at

$$\delta(t) \approx \frac{B}{3T}\left(1 - e^{-t/4\tau_0}\right). \tag{S17}$$

Finally, the reconfiguration current is given by

$$J_{\text{rec}} = \frac{dp_+}{dt} - \frac{dp_-}{dt} \approx 2\frac{d\delta}{dt} \approx \frac{B}{6\tau_0 T}e^{-t/4\tau_0}. \tag{S18}$$

To summarise, if we start in thermodynamic equilibrium in zero field and then suddenly turn on a field satisfying $0 < B \ll T$ at time $t = 0$, the resulting currents are

$$J_{\text{jagged}} = \frac{B}{12\tau_0 T} + \frac{B}{6\tau_0 T}e^{-t/4\tau_0} + O((B/T)^2) \tag{S19}$$

and

$$J_{\text{dense}} = \frac{5B}{18\tau_0 T} + O((B/T)^2) \tag{S20}$$

Both in the jagged and dense case, the currents are linear in $B/T$. These final two equations are plotted in Fig. S5.

**Relation to spin ice**

In spin ice simulations or experiments we typically measure the magnetisation/magnetic flux, rather than the monopole current directly. The closest analogue to the magnetisation in the ladders is given by integrating the total current:

$$M_{\text{jagged}}(t) = \int_0^t dt'\, J_{\text{jagged}} = \frac{B}{12T}\frac{t}{\tau_0} + \frac{4B}{6T}(1 - e^{-t/4\tau_0}) \tag{S21}$$

and

$$M_{\text{dense}}(t) = \int_0^t dt'\, J_{\text{dense}} = \frac{5B}{18T}\frac{t}{\tau_0}. \tag{S22}$$

Unlike for spin ice, this effective magnetisation can keep growing indefinitely – the ladders support DC currents. Despite this important difference, the behaviour is qualitatively similar at short times (see Fig. S6).

Notice that the reconfiguration current identified in our jagged ladder is a single particle effect. It decays on a time scale $4\tau_0$ which we generally do not expect to be affected significantly by the presence of other charges (at sufficiently low densities at least), and as such it is density-independent. We expect this to be the case also for the reconfiguration current decay time scale $\tau_R$ in spin ice, discussed in the main text. The polarisation current in spin ice, on the other hand decays as the system becomes polarised (a phenomenon not captured by our ladder models), and this occurs on a timescale which depends on the strength of the polarising monopole currents (5). It is thus expected that this time scale, which we call $\tau_P$ in the main text, depends on the density of monopoles and thus is temperature-dependent. The considerations here thus provide a compelling explanation for our finding that the response of spin ice to a sudden applied external field is governed by a temperature-independent, shorter time scale $\tau_R$ (which we relate to the experimentally measured $\tau_1$) and a temperature-dependent, longer time scale $\tau_P$ (which we relate to the experimentally measured $\tau_2$).

**Oscillating field**

We now turn to the case of an applied field that oscillates as $B = B_0\sin(2\pi f t)$. We consider sufficiently weak fields to assume that the fluctuations in $p_0$ are small. To leading order, the steady state current is then $J_{ss} = \frac{B_0}{12\tau_0 T}\sin(2\pi f t)$, where we have approximated $p_0$ as $1/3$ and expanded the exponential to first order in $B/T$. It is in phase with the applied field, which is what one intuitively expects. These results are indeed confirmed by our numerics in Fig. S7.

A similar argument works for $J_{rec}$ at high frequency. When the period of the applied field is smaller than the time scale on which the probabilities can change, $\tau_0$, the probabilities $p_+$ and $p_-$ do not have time to vary significantly before the field reverses, and we can again approximate $p_- \approx 1/3$ and $p_+ \approx 1/3$, from which one finds $J_{rec} \approx \frac{2B_0}{12\tau_0 T}\sin(2\pi f t)$,. Again, this is confirmed by our numerics in Fig. S7.

In the limit of low frequency, one can instead assume that $p_-$ and $p_+$ are in equilibrium, and take the values given by Eq. (S4). The reconfiguration current is then given by

$$J_{rec}^{(eq)} = \left[\frac{dp_+^{(eq)}}{dt} - \frac{dp_-^{(eq)}}{dt}\right]$$

$$= \left[\frac{dp_+^{(eq)}}{dB}\frac{dB}{dt} - \frac{dp_-^{(eq)}}{dB}\frac{dB}{dt}\right]$$

$$\approx \frac{4\pi f B_0}{3T}\cos(2\pi f t),\quad (S23)$$

where we have expanded to first order in $B_0/T$. This is consistent with the phase shift being $\pi/2$ for $J_{rec}$ at low frequency (see Fig. S7, top panel). If we plot it on a log-log scale, we see that the amplitude of $J_{rec}$ follows this form (see Fig. S7, bottom panel). In fact, the amplitude follows this equilibrium approximation essentially all the way until the high frequency plateau that was explained in the previous paragraph.

The physics we want to understand with this model is the peak in the phase shift of the total current. It is now clear that this appears at the crossover where the reconfiguration current goes from being $\pi/2$ out of phase, but very small, to being the dominant contribution to the current, but in phase. At this crossover there is a regime where the reconfiguration current is of comparable size to the steady state current, while still out of phase with the applied field. This results in the observed phase shift of the total current across an intermediate range of frequencies.

One can numerically integrate Eq. (S2) and Eq. (S3) and use them to obtain $J_{jagged}$, $J_{ss}$, and $J_{rec}$ for a range of frequencies, $f$ (see Fig. S8) and Fig. S7 for the extracted phases and amplitudes of the currents). The phase of the total current shows a peak at intermediate frequencies, similar to the behaviour observed for spin ice.

Coming back to spin ice, the main difference with respect to the effective ladder models is that spin ice does not support a steady state current. This is why the magnetisation does not go back to being $\pi/2$ out of phase with the applied field at low frequency. Nevertheless, the analogy to the jagged ladder provides a useful simple scenario to develop insight into the behaviour of charged particles moving under the influence of a field on a lattice with termini in bSM spin ice.

**Finite density of bonds between the legs**

One can easily extend the results for the jagged ladder to the more general case where the bonds connecting central sites to sites on the upper or lower ladder are removed at random with probability $1-q$. Eq. (S2) and (S3) become

$$\frac{dp_-}{dt} = \frac{1}{4\tau_0}[q(1-p_- -p_+)Q_- - p_- Q_+] \tag{S24}$$

and

$$\frac{dp_+}{dt} = \frac{1}{4\tau_0}[q(1-p_- -p_+)Q_+ - p_+ Q_-], \tag{S25}$$

and the equilibrium values are given by

$$p_\pm^{(\text{eq})} = \frac{e^{\pm B/T}}{q^{-1}+e^{-B/T}+e^{B/T}}. \tag{S26}$$

Following the same steps as in the $q = 1$ case, leading to Eq. (S19), one can show that the current on the jagged ladder after a constant weak field is suddenly applied at time $t = 0$ is

$$J_{\text{jagged}} = \frac{1}{1+2q}\frac{B}{4\tau_0 T} + \frac{1}{q^{-1}+2}\frac{B}{2\tau_0 T}e^{-t/4\tau_0} + O((B/T)^2). \tag{S27}$$

The relative contribution of the steady state (first term) and reconfiguration (second term) currents for $t \to 0$ is thus

$$\frac{J_{\text{rec}}(t=0)}{J_{ss}(t=0)} = 2q, \tag{S28}$$

and for very small $q$, the contributions of the upper and lower legs eventually become negligible.

In the case of an oscillating field, the steady state current is simply re-scaled to account for the new value of $p_0$ – giving $J_{ss} = \frac{B_0}{4(1+2q)\tau_0 T}\sin(2\pi f t)$, Similarly, using the $q$-dependent zero-field values of $p_\pm$ we find that $J_{\text{rec}} \approx \frac{2B_0}{4(q^{-1}+2)\tau_0 T}\sin(2\pi f t)$, at high frequency. In the low frequency limit we can reproduce the steps given in Eq. (S23) to obtain

$$J_{\text{rec}} \approx \frac{12\pi f B_0}{(q^{-1}+2)^2 T}\cos(2\pi f t). \tag{S29}$$

The contribution of the upper and lower legs again become negligible for very small $q$, but the general behaviour is otherwise preserved in presence of random bond dilution ($q < 1$) – namely, the results in this section are robust and not limited to fine-tuned regular ladders.

## (IV) Monopole Current and Noise Spectrometer

**Apparatus design**
All the experiments reported in this paper were carried out with the same instrumental setup in the custom-built 1 Kelvin (1K) cryostat. The whole 1K refrigerator is mounted on the isolated ultra-low vibration keel-slab in an acoustically isolated laboratory, with the electronics and data systems in the adjacent control room. The schematic illustration of the spectrometer is shown in Figure 1A. The spectrometer assembly consists of a superconductive pickup coil wound directly on the sample holder and an enclosing superconductive excitation coil. The cylindrical sample holder is designed with a concentric hole of 1.6 mm diameter and length 5 mm, in order to encapsulate the rod-shaped sample. The pickup coil which is made from a single wire consists of two in-series counter-wound NbTi coils with 10 turns each. The excitation coil is wound with NbTi for 537 turns and 90 mm in length. The mid-point of pickup coil is at the center of the excitation coil. The overall spectrometer assembly is mounted below the refrigerator and is shielded by a Nb shield which has an inner diameter of 47.9 mm and 2 mm in thickness. The pickup coil, excitation coil, and Nb shield are concentrically aligned. The SQUID is mounted next to 1K pot to maintain the low SQUID temperature while varying the sample temperature during the experiments. The QD SP550 SQUID is shielded separately with another Nb shield provided by the manufacturer (Quantum Design) and the input inductance $L_i$ is reported to be 1.82 µH. To deliver the detected flux signal, the pickup coil enters the SQUID externally and is shielded inside the superconducting Pb alloy tube.

**Experimental operation and circuit diagram**

A typical experimental protocol starts from cooling the sample down to the target temperature with the range from 1.7 K to 4.5 K. The $Dy_2Ti_2O_7$ sample is stabilized at the target temperature for 10 minutes with the stability of 5 mK before measurements. Once the temperature is stable, a specific waveform of field $B(t)$ is applied along the long axis of the rod-shaped sample in the crystal [351] direction based on different experiments. To control the magnetic field, we use the voltage output of Keysight's 33500B waveform generator in series of a 1 kΩ resistor and the excitation coil circuitry. Fig. S9 shows the overall circuit diagram of the experiment. The pickup coil consists of three inductors labeled with $L_p$ and $L_i$. The sample generates the flux $\Phi_p$ threading through one of the pickup coil. The SQUID records the $\Phi_S(t)$ and converts it into the voltage signal in the time domain $V(t)$. $\Phi_p$ and $\Phi_S$ are connected by the coupling constant $\alpha$, while the SQUID output voltage is proportional to $\Phi_p$ with a transfer function $\gamma$ in Eq. (2). The detail of the two constant is described in the later section. The output voltage is then measured by a 16-bit ADC for the time-domain, and a lock-in amplifier for the frequency domain, measurements. The time-dependent and frequency-dependent measurements were performed for the same sample in the same temperature range.

**Empty coil calibration**

The flux that is collected by the pickup coil ($\Phi_\text{p}$) and the flux that SQUID senses ($\Phi_s$) are linked with a coupling constant $\alpha$.
$$\Phi_\text{S} = \alpha \Phi_\text{p} \tag{S30}$$
where $\alpha$ can be derived by
$$\alpha = \frac{M_\text{i}}{2L_\text{p} + L_\text{i}} \tag{S31}$$
$M_i$ is the mutual inductance between SQUID and pickup coil with $M_\text{i} \approx 9.84 \times 10^{-3}$ µH. $2L_\text{p}$ is 1.7 µH measured by the LCR meter at room temperature. The measured parameters yield $\alpha \sim 2.8 \times 10^{-3}$. For clarity, the flux in all the figures of the manuscript is reported in $\Phi_\text{s}$. The flux at the SQUID $\Phi(t)$ is linearly proportional to the voltage output, it can be written as
$$\Phi_\text{S}(t) = V_\text{S}(t)/\gamma \tag{S32}$$
For this SQUID the transfer function $\gamma$ is known to be 0.75 V/$\varphi_0$ for the highest sensitivity.

Fig. S10 demonstrates the cancellation of external field of the two-counter wound pickup coil. The measured flux with the applied field is shown as the blue curve, while the red curve is the expected flux if there is no compensation coil. It is estimated that 98.7% of the flux is compensated leaving only 1.3% of the applied flux detected.

## (V) Dy$_2$Ti$_2$O$_7$ sample synthesis

The single crystal samples of Dy$_2$Ti$_2$O$_7$ were grown using floating zone method. High purity Dy$_2$O$_3$ (99.99%), and TiO$_2$ (99.99%) were mixed and heated to 1400° C for 40 hours and then again for 12 hours after intermediate griding. This powder was then packed into a rod which was sintered at 1400° C for 12 hrs. A longer piece of the sintered rod was used as a feed rod and smaller piece was used as seed in the floating zone furnace. The growth was carried out in a 0.4 MPa oxygen pressure at 4 mm/hr using a two mirror NEC furnace where the seed and feed rods were counter rotated at 30 rpm. The Dy$_2$Ti$_2$O$_7$ crystal is then cut into rod-shaped sample with the geometry of 1.3 mm x 1.3 mm x 6.5 mm with the long axis along the [351] direction.

## (VI) Time dependent flux transient measurement

**Data acquisition procedure**
To measure the transient of monopole current in both positive and negative field direction, the excitation field is designed to be a sequence of square waves in the two directions (green,

Fig. 2A). The field strength of the excitation field is about 55 nT and remains for 2.5 ms, followed by a zero field period for 10 ms. The overall duration of measurement is normally over 2 minutes. SQUID is set at the highest sensitivity. Simultaneously, the 16 bit-ADC records the output voltage from the SQUID electronics at the sample rate of 1MSa/s and for typically over 1000 transients of excitation at each measured temperature. Experimentally, the SQUID has a flux feedback equilibration time of less than 200 μs. Therefore, all the time series analysis excludes the first 200 μs.

**Averaging of flux transients**
To further increase the signal-to-noise ratio (SNR), the measured flux $\Phi(t)$ is averaged for all the transients. Fig. S11a-c shows schematic diagram of the averaging process. We first cut the corresponding flux $\Phi(t)$ into $k$ sections, and $t$ = 0 is defined as the time when $B$ field is turn on at each section.

$$\Phi(t_j) = \begin{Bmatrix} \Phi_1(t_j) \\ \Phi_2(t_j) \\ ... \\ \Phi_k(t_j) \end{Bmatrix} \tag{S33}$$

The flux response $\Phi(t)$ is later averaged by k sections as
$$\langle \Phi(t) \rangle = \frac{1}{k} \sum_{i=1}^{k} \Phi_i(t_j) \tag{S34}$$
An example of the average flux response can be seen in Fig. S11b, where the random noise is strongly suppressed. Experimentally, the correct flux response of the sample is calculated from the difference of the measured flux from any offset due to the environmental background. Thus, to remove the offset, the flux in the positive field direction is subtracted by the offset. Also, the sign of flux flips to compare with a single time scale exponential decay.
$$\Phi^+ = -(\langle \Phi(t) \rangle - \langle \Phi(t_j = 1.8 \text{ ms}) \rangle) \tag{S35}$$
Similarly, in the negative field direction, flux response is calculated as
$$\Phi^- = \langle \Phi(t) \rangle - \langle \Phi(t_j = 1.8 \text{ ms}) \rangle \tag{S36}$$

**Separation of fast and slow monopole current**
In the two-time-scale model, the flux consists of the contribution of two independent channels. For example, in the positive field direction
$$\Phi^+(t,T) = \Phi_1^+(t,T) + \Phi_2^+(t,T) \tag{S37}$$

In the long-time limit, $\Phi_1$ would approach zero. Therefore, the fitting range for $\Phi_2$ is chosen to be 600 μs to 1000 μs where the fast current contribution $\Phi_1$ is negligible and the flux signal is above noise level. We found that $\Phi_2$ is best described by fitting $\log \Phi_2$ with a linear function with the coefficient $\tau_2$ representing the relaxation time
$$\log \Phi_2(t, T) = C(T) - t/\tau_2(T) \tag{S38}$$
Fig. S12a and 4b illustrate the example of the two-current fit. The quality of the fit is shown in the Fig. S12c. The flux contributed from the fast current $\Phi_1$ is thereby calculated by subtracting $\Phi_2$ from the slow current
$$\Phi_1(t, T) = \Phi(t, T) - \Phi_2(t, T) \tag{S39}$$
Similarly, the time scale $\tau_1$ for the fast current is extracted by fitting $\log \Phi_1(t, T)$
$$\log \Phi_1(t, T) = A(T) - t/\tau_1(T) \tag{S40}$$

## (VII) Frequency dependent monopole conductivity measurement

### Data acquisition procedure
Magnetic monopole AC conductivity was measured by a lock-in amplifier which picks up the in-phase and out-of-phase components from the voltage output of the SQUID. The SQUID is set at the sensitivity of range 500. The waveform generator produces an AC sinusoidal drive to the primary coil corresponding to a field of 11 nT in the typical frequency range of 10 Hz to 6000 Hz. The reference of the lock-in amplifier is set by measuring the actual current using the voltage drop across the 1 kΩ resistor in series (see Fig. S9b). This is to compensate the phase change due to the LR circuit of the primary coil. The zero phase is set by performing a calibration experiment without a sample. The time constant of the lock-in amplifier was set at 1 s for frequency below 100 Hz and 100 ms for frequencies above 100Hz. The low pass filter was set at 18 dB/oct for all frequency. The sensitivity of the lock-in amplifier was set to be 5 mV for all the frequencies.

### Complex conductivity and loss angle of monopole current
For an input signal $V(t)$, the lock-in amplifier (SR830) output X and Y channels are defined
$$V(t) = V_X \sin(2\pi f t) + V_Y \cos(2\pi f t) \tag{S41}$$
Therefore, the real and imaginary part of the AC conductivity of magnetic monopoles can be derived from the measured in-phase $\Phi_X$ and out-of-phase $\Phi_Y$ of the modulated flux.
$$\Phi(t) = \Phi_X \sin(2\pi f t) + \Phi_Y \cos(2\pi f t) \tag{S42}$$
The monopole current is proportional to time derivative of flux by $J(t) = \dot{\Phi}(t)/\mu_0$. From Eq. (S42), we can derive

$$\dot{\Phi}(t) = -2\pi f \Phi_Y \sin(2\pi f t) + 2\pi f \Phi_X \cos(2\pi f t) \tag{S43}$$

Therefore, the real part of conductivity $J_X$ is

$$J_X(f) = \frac{-2\pi f \Phi_Y(f)}{\mu_0} \tag{S44}$$

While the imaginary part can be expressed as

$$J_Y(f) = \frac{2\pi f \Phi_X(f)}{\mu_0} \tag{S45}$$

The loss angle is defined as

$$\theta(f) \equiv \arctan\left(\frac{J_Y(f)}{J_X(f)}\right) = \arctan\left(-\frac{\Phi_X(f)}{\Phi_Y(f)}\right) \tag{S46}$$

To clearly visualize the inflection point in the frequency-dependent loss angle, we performed a numerical derivative over frequency by

$$\frac{d\theta_i}{df} = \frac{\theta_{i+1} - \theta_i}{f_{i+1} - f_i} \tag{S47}$$

## (VIII) Noise analysis

To calculate the variance of the two current regimes, the flux deviation $\delta\Phi$ of each section to the averaged flux response is first obtained by

$$\delta\Phi_i(t_j) = \Phi_i(t_j) - \langle\Phi(t)\rangle \tag{S48}$$

where the $\langle\Phi(t)\rangle$ is the averaged transient response acquired from Eq. (S34). Next, the variance is the average of the square of the deviation.

$$\sigma^2(t_j) = \frac{1}{k}\sum_{i=1}^{k} \delta\Phi_i^2(t_j) \tag{S49}$$

To further exclude the random noise, the time-dependent variance is averaged in time for a bin size of $t_{bin}$ so that

$$\sigma_n^2(t_j) = \sum_{t_j = nt_{shift}}^{nt_{shift} + t_{bin}} \sigma^2(t_j) \quad (n = 0, 1, 2 \dots) \tag{S50}$$

Fig. S13a shows the flux variance over time at the 2.2 K with the $t_{bin}$= 50 μs, $t_{shift}$= 50 μs. Clearly, a burst of the noise can be seen from the first measurable 100 μs, from 200 μs to 300 μs after the magnetic field is switched on. Fig. S13b shows flux variance over time for different temperatures, each averaged with the $t_{bin}$= 100 μs, $t_{shift}$= 50 μs. The increase of the flux noise is commonly seen for the measured temperature range. For best representing the specific noise power when the fast current exists, the time series variance $\sigma_1^2$ for the fast current is further averaged within the first 100 μs.

$$\sigma_1^2 = \frac{1}{100}\sum_{t_j = 200\ \mu s}^{300\ \mu s} \sigma^2(t_j) \tag{S51}$$

On the other hand, the variance representing the absence of the fast current is calculated in the time interval from 600 μs to 800 μs after the magnetic field is switched on.

$$\sigma_2^2 = \frac{1}{200}\sum_{t_j=600\,\mu s}^{800\,\mu s} \sigma^2(t_j) \tag{S52}$$

The 600 µs is chosen as the fast current completely ceases and therefore the magnetic flux noise is dominated by the generation-recombination noise. The averaging process increases the signal-to-noise ratio and thereby minimizes the systematic error in time measurement.

## (IX) Spin ice dynamics beyond the standard model

Within the bSM model (*23*), spins in spin ice flip at one of two different characteristic time scales, $\tau_{\text{fast}}$ and $\tau_{\text{slow}}$, depending on whether there is a non-vanishing or vanishing transverse magnetic field across them. Note that spin flips that create or annihilate monopole pairs always occur with the fast time scale. However, as the creation of a pair carries a large energy penalty, these events are nonetheless rare at low temperatures and magnetisation dynamics becomes dominated by monopole motion. Although the application of a magnetic field does change the cost of generating a monopole pair, this change is negligible at the field strengths considered here.

One does have to ask under what conditions the bSM is applicable in presence of external fields. The typical magnitude of the internal transverse fields that facilitates spin flips is approximately 0.05 T and 0.5 T, for the slow and fast spins respectively (*22*). If the applied field is $\lesssim$ 0.05 T, the bSM should still be a reasonable approximation of the magnetisation dynamics in the material.

In our simulations we have used an effectively infinite $\tau_{\text{slow}}$. The theoretical prediction from microscopic single ion calculations is that $\tau_{\text{slow}}/\tau_{\text{fast}} \sim 10^4$ (*22*), and, as in Ref. (*23*), we find that the results are independent of the exact choice of $\tau_{\text{slow}}$ if $\tau_{\text{slow}}/\tau_{\text{fast}} \gtrsim 10^3$ (see Fig. S14). This also demonstrates that the two current regimes discussed in the main text are not trivially related to the two time scales on which spins can flip, as one might have naively expected, and the underlying connection is more subtle – as we uncover in this work.

The development of the bSM dynamics was motivated by the observation of anomalous magnetic noise in $Dy_2Ti_2O_7$ (*20,21*). The anomalous noise is indeed explained by an important consequence of the bSM – the combination of dynamical and energetic constraints constrict monopoles to move on sparse emergent clusters. These clusters appear fractal on small and intermediate length scales, and it is this fractal nature that manifests itself in the anomalous exponent of the magnetic noise. Note that, in this picture, monopoles are sparse and move each on their own cluster, formed by the sites that the monopoles can access within the remit of the ice rules and fast-flipping spins.

The phenomena resulting from the dynamical field-driven protocols investigated in this work are another intriguing and exciting consequence of the peculiar internal field distribution and local time scales of the bSM dynamics. However, they do not derive from the fractal nature of the clusters on which sparse monopoles move. Instead, their origin is rooted in the presence of local dead-ends, or termini, that directly affect monopole response when driven by an applied magnetic field. This is most crisply illustrated by the simple 'jagged ladder' model that we present below, and that we feel captures the essence of the dichotomous response discussed in the main text.

## (X) Additional numerical results

The best available modelling of $Dy_2Ti_2O_7$ indicates that its magnetic behaviour is described by the Hamiltonian $\mathcal{H}_{\text{OP}}$ (see Methods). This Hamiltonian includes long-ranged dipolar interactions, making it computationally expensive to simulate for larger systems. The much simpler Hamiltonian

$$\mathcal{H}_{\text{NN}} = -J_{\text{eff}} \sum_{\langle i,j \rangle} \vec{S}_i \cdot \vec{S}_j \,, \tag{S53}$$

including only nearest neighbour interactions, reproduces the dynamical and thermodynamic behaviour of $\mathcal{H}_{\text{OP}}$ to a good approximation. We use it here both for convenience of accessing longer times and better statistics, as well as to test the dependence of qualitative features in the behaviour of the system as a function of system size.

The relevant parameters for $\mathcal{H}_{\text{NN}}$ simulations in presence of a field $B$ at temperature $T$ are the ratios $J_{\text{eff}}/T$ and $B/T$. We then map to realistic physical temperature units by measuring the density of magnetic monopoles in thermodynamic equilibrium at zero field, and mapping the temperature used in the nearest-neighbour simulation to the temperature that would give the same monopole density for $\mathcal{H}_{\text{OP}}$. With $\mathcal{H}_{\text{NN}}$ we can access larger systems, and in this Supplementary Information all results presented are for $\mathcal{H}_{\text{NN}}$ and with a system of 128,000 spins.

In order to verify that the system is in the linear response regime in our simulations we have measured the response to a sudden applied field for field strengths between 7.5 mT and 30 mT (see Fig. S15). The magnetisation response is clearly linear in the field strength both in the short-time and long-time limits. This shows that both the reconfiguration and polarisation currents are proportional to the magnitude of the applied field. We have also measured the response to a sudden applied field applied along different lattice directions, and found that this has no effect on the observed response functions (see Fig. S16).

The only free parameter in our theory is the fast-flipping timescale $\tau_{fast}$. In Ref. (*23*), fits to experimental magnetic noise data gave $\tau_{fast} \approx 85$ μs. In the present work, we use this same value without any further fitting. We extract the current decay times $\tau_R$ and $\tau_P$, related to the reconfiguration and polarization currents respectively, from simulations; in Fig. S17 we compare them to the time scales $\tau_1$ and $\tau_2$ found to govern the decay of the currents in our experiments. We naturally relate the faster-decaying of the two currents to the reconfiguration of monopoles, and the agreement between $\tau_1$ and $\tau_R$ is excellent. $\tau_P$ displays a stronger temperature dependence than $\tau_2$, but the two are of similar magnitude. Note that $\tau_R \approx \tau_{fast}$; the local reconfiguration of monopoles happens on the order of one monopole move.

## (XI) Monopole Density and Transport Dynamics at High Temperatures

The bSM model has previously primarily been considered at temperatures around 1 K (*22,23*). In this section, we therefore discuss and motivate out use of the bSM model at temperatures up to 4 K.

Firstly, the underlying physical arguments in ref. (*22*) that motivate the bSM model is that the transverse field distribution across spins takes on a bimodal form, with one group of spins being subject to a transverse field of approximate magnitude 0.03 T, and the remaining spins being subject to a transverse field of approximate magnitude 0.45 T. (It should be noted that these computed transverse fields include contributions from monopoles, in the sense that the field coming from a monopole is simply the net field coming from the spins of a tetrahedron that hosts a monopole). Importantly, the dominant contribution to the transverse field is given by the six nearest-neighbours around each spin – with the transverse field from spins further away only contributing some broadening of the sharp peaks from the nearest-neighbour contribution.

Secondly, it should be noted that the flipping of spins not corresponding to monopole moves is also governed by the local transverse field. However, for such spin flips the local transverse field is significantly larger than the ~0.03 T that some monopole-hopping spins experience, because the six nearest-neighbour spins almost never have a vanishing transverse field in calculation. The only exception is for a spin sitting between a single monopole and a double monopole of the same sign, which is a very rare occurrence that we do account for in our simulations. At the level of the bSM model, moves that do not hop monopoles are therefore treated as if they occur on the fast time scale (with the one exception just explained). Certainly, the distribution of transverse fields for these spins that do not hop monopoles is broader than the distribution for spins that do hop monopoles, and the main bSM approximation of only considering two flipping time scales is therefore less precise. This will primarily come into effect at higher temperatures, where spin flips that carry a large energy penalty are more common. The most important impact of the bSM model, namely that some spins are essentially blocked from flipping due to the very small transverse field they experience, is however still accurately accounted for.

According to the Debye-Hückel theory for spin ice (*29*), the monopole density per tetrahedron at equilibrium and in the absence of any external magnetic field is given by

$$\rho_0 = \frac{2\exp(-\Delta/T)}{1+2\exp(-\Delta/T)}. \tag{S54}$$

where the gap energy is Δ = 4.7 K. From this expression, we compute that approximately 15% of tetrahedra in $Dy_2Ti_2O_7$ hosts a monopole at a temperature of 2 K, whereas that number is approximately 40 % at a temperature of 4 K. It is thus not correct to treat transport in the independent monopole limit at 4 K. As the discussion above hopefully has made clear, our simulations are not reliant on this limit. Our phenomenological interpretation of the results is, on the other hand, reliant on monopoles being the dominant drivers of magnetic response. This does not rely on the monopole density being low, as each monopole only needs to move a fraction of a step or at most a couple of steps to relax the system for these weak applied fields. At some level, this explains why dichotomous behaviour persists to relatively high temperatures. The observation of dichotomous transport properties up to 4 K is however still surprising, and indicates that the impact of the transverse field distributions on the response continues to play a key role. At 4 K our Monte Carlo simulations do not observe dichotomous properties, suggesting that the bSM model may no longer be an accurate description of the system

It is also useful to compare the field generated by the neighboring monopole and the transverse field imposed by the slow/fast spin configuration. The magnetic field $B_m$ of a monopole impose onto the neighboring monopole without screening can be obtained from the magnetic Coulomb law that

$$\vec{B}_m = \frac{\mu_0}{4\pi}\frac{m^*}{r^2}\hat{r}. \tag{S55}$$

By using the lattice distance of $r_d = 4.34$ Å, it is estimated that for a monopole directly next to another monopole in the neighboring tetrahedron, this field is around 0.2 T. While the fast spin-flipping configuration calculated (*22*) is approximately 0.45 T, and the slow spin-flipping configuration is approximately 0.03 T.

## (XII) Estimation of Demagnetization factor

The demagnetization factor for a sample in rod geometry with the field aligned along the rod approaches zero in the limit of infinite length. However, in practice, a finite size rod-shaped sample has a demagnetization field $B_d$ of

$$\vec{B}_d = -N(\vec{r})\mu_0\vec{M}. \tag{S56}$$

where demagnetization factor is $N(\vec{r})$ and is position dependent. $N(\vec{r})$ for a cuboid with a dimension of 2a × 2b × 2c can be expressed as[28]

$$N_{ii}(\vec{r}) = \frac{1}{4\pi}\sum_{\alpha=\pm 1}\sum_{\beta=\pm 1}\sum_{\gamma=\pm 1}\tan^{-1}(f_i(\alpha x, \beta y, \gamma z)); \; i = x, y, z \tag{S57}$$

where

$$f_x(x,y,z) = \frac{(b-y)((c-z)}{(a-x)\sqrt{[(a-x)^2+(b-y)^2+(c-z)^2]}} \tag{S58}$$

$$f_y(x,y,z) = \frac{(a-x)((c-z)}{(b-y)\sqrt{[(a-x)^2+(b-y)^2+(c-z)^2]}} \tag{S59}$$

$$f_z(x,y,z) = \frac{(b-y)((a-x)}{(c-z)\sqrt{[(a-x)^2+(b-y)^2+(c-z)^2]}} \tag{S60}$$

We argue that for the flux threading the superconducting coil which winds at the centre of the DTO sample, $N_{zz}(0,0,0)$ contributes to the measurement the most. By using the dimension of 2a = 1.3 mm, 2b = 1.3 mm, 2c = 6.5 mm, this yields an effective demagnetization factor of

$$N_{zz}(0,0,0) \sim 0.0113 \tag{S61}$$

Therefore, the estimated demagnetization factor is roughly 1%. The observation of two timescale monopole transport and the derived relaxation time are not affected by the demagnetization field. Quantitatively, a small fraction of 1% of the measured flux could be accounted for the demagnetization factor.

## (XIII) Relaxation time constants

It could be helpful to relate the relaxation time constants derived in this work to the relaxation time constants measured in the literatures. Figure S18a shows a combined plot of relaxation time constants from AC susceptibility (ref. 7,10,12), correlation (ref. 11) and DC magnetization measurements (ref. 13,14). In this work, we report reconfiguration current time constant $\tau_1$, which we propose corresponds to the quantum mechanical spin flipping time scale $\sim$ 90 $\mu$s. On the other hand, the polarisation current time constant $\tau_2$ is a slow-decaying collective magnetization time scale. In our measured temperature range, $\tau_1$ roughly remains constant while $\tau_2$ increases by around 20%. Compared to the literatures, our temperature range is measured where the relaxation time exhibit small changes in the temperature dependency.

Note that the extracted slow-decaying time constants may have quantitative discrepancy from the AC susceptibility measurements. Besides the fact that the relaxation function is very different, one possible explanation is the different magnetic field strength used for the measurements. In this work, a magnetic field of 55 nT is used, and therefore we measure the perturbation close to the zero field limit. On the other hand, in the case of higher magnetic field, DTO can be polarized more and microscopically has the lower monopole density in the magnetized state. The relaxation process is thereby quantitatively different from the relaxation process in the normal state near zero field. More specifically, we have performed the time domain flux measurement as Figure 2A with higher field (550 nT). As shown in Figure S18b, the extracted $\tau_2$ has a more rapid change with the temperature. While the fast spin flipping time constant $\tau_1$ remains constant. This result implies that the microscopic monopole dynamics may be altered by the different magnetic field strengths. We leave further investigation of this potential field dependence as an avenue for future work.

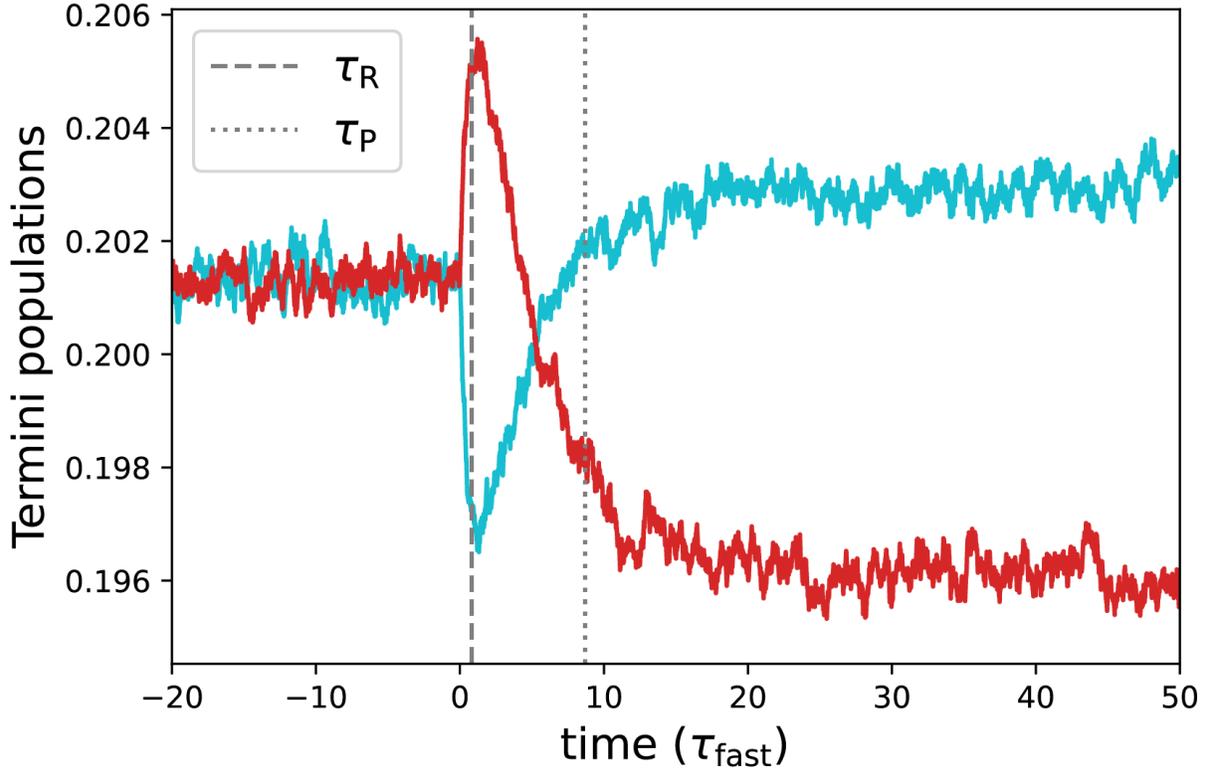

**Fig. S1**: The proportion of monopoles that are in a positive terminus ($\gamma_+$; red) and in a negative terminus ($\gamma_-$; cyan) is shown for $T/J_{\text{eff}} = 0.53$ (corresponding to approximately 2.2 K). Initially, there is no field applied and $\gamma_+ = \gamma_-$. At time $t = 0$ a field of magnitude 15 mT is applied along the [100] direction, causing monopoles to move and $\gamma_+$ ($\gamma_-$) to increase (decrease). After a time on the order of $\tau_R$ (vertical dashed line), the reconfiguration of monopoles is complete. In a polarised state (positive net magnetisation in the [100] direction), configurations with monopoles in negative (positive) termini are energetically (dis)favoured and we observe a corresponding decrease in $\gamma_+$ and increase in $\gamma_-$, on a time scale $\tau_P$ (vertical dotted line). Data from simulations with $\mathcal{H}_{\text{NN}}$ and a 128,000 spin system.

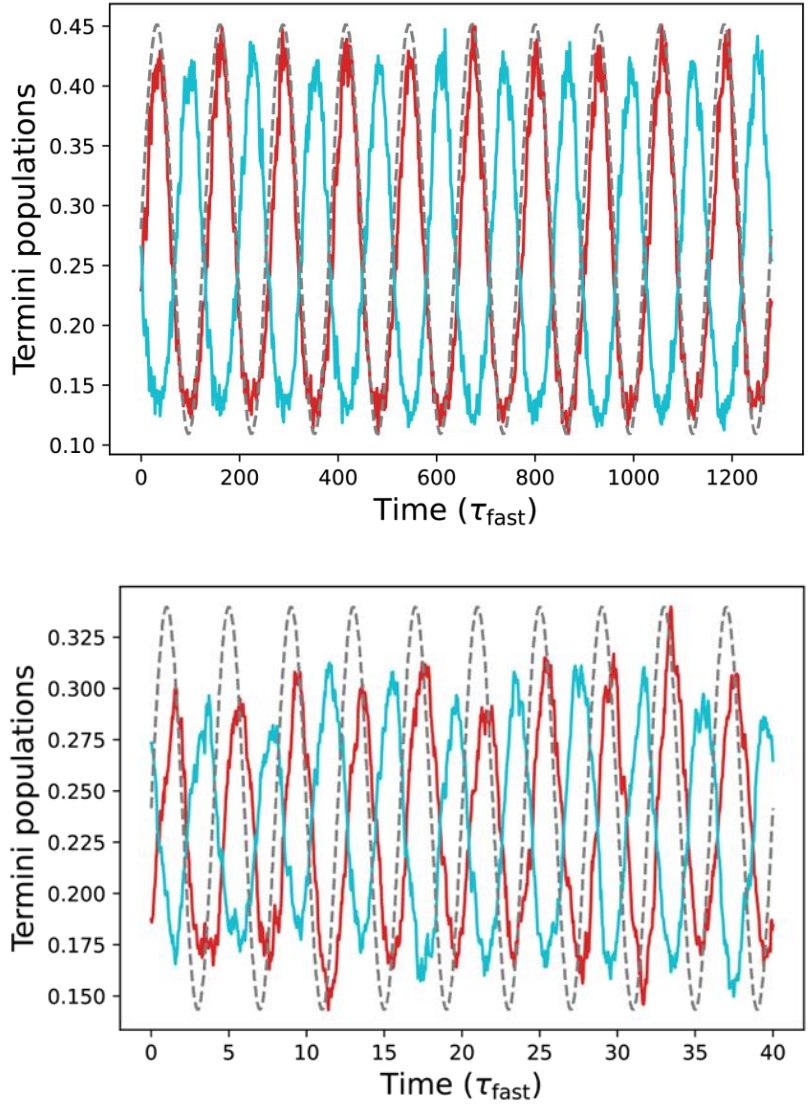

**Fig. S2**: Examples of how the termini populations γ₋ (cyan) and γ₊ (red) change in an applied field $H(t) = H_0\sin(2\pi f t)$ at different driving frequencies $f = 7.8 \times 10^{-3}\tau_{fast}^{-1}$ (upper panel), and $f = 2.5 \times 10^{-1}\tau_{fast}^{-1}$ (lower panel). The modulation of the applied field is shown by the dashed grey line. For frequencies $f \ll \tau_R^{-1}$, the termini populations oscillate in phase with the applied field. $\gamma_-$ and $\gamma_+$ are effectively in equilibrium with the applied field, and their change generates the reconfiguration current that gives bSM spin ice the characteristic intermediate-frequency bump in the loss angle. At high frequency the field changes too rapidly for the termini populations to keep up, and, as is the case for the magnetisation, $\gamma_-$ and $\gamma_+$ lag behind the applied field by approximately π/2. Data from simulations with $\mathcal{H}_{\text{NN}}$ and a 128,000 spin system.

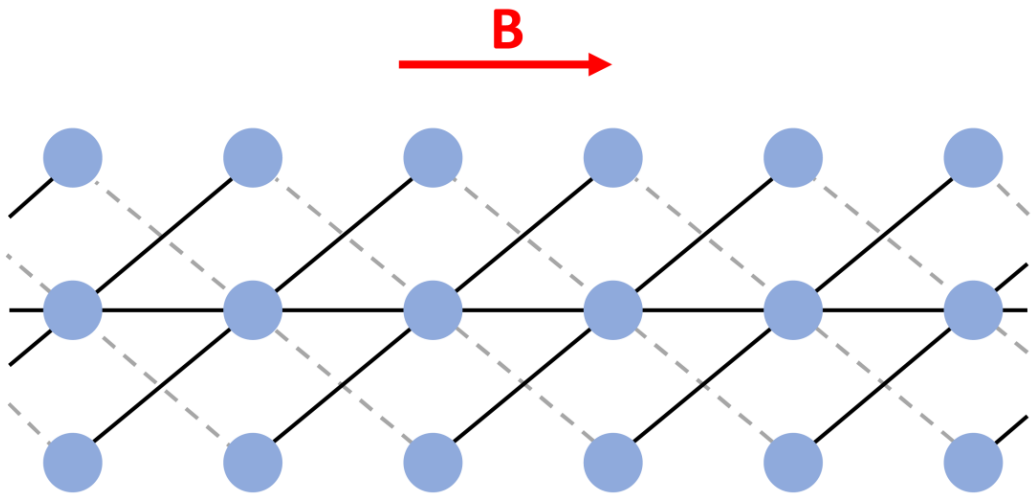

**Fig. S3**: The geometry of the three-leg ladders that we are considering: the jagged ladder, with connectivity drawn by solid black lines; and the dense ladder, where in addition to the solid black lines, we also allow the connections between sites drawn by dashed grey lines. The arrow indicates the chosen positive direction for the applied field (without loss of generality).

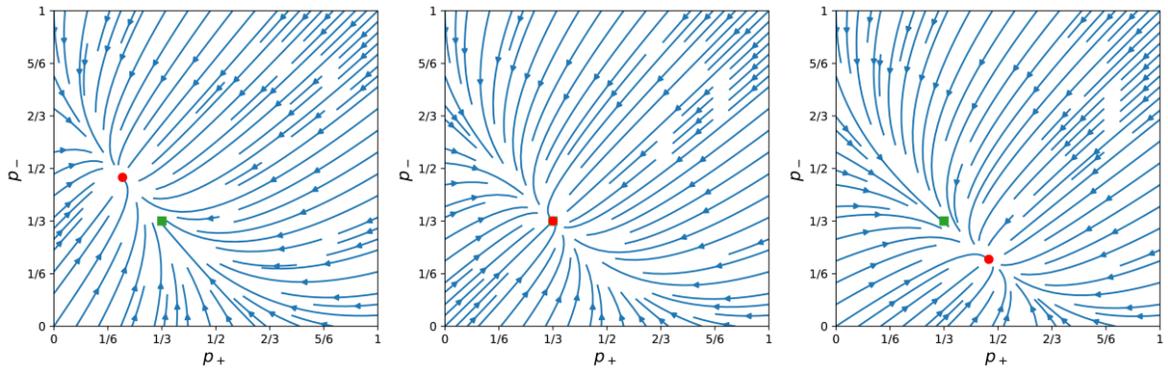

**Fig. S4**: Diagrams showing the flow of $p_+$ and $p_-$, computed from Eq. (S2) and (S3), for $B/T$ = −0.4 (left panel), $B/T$ = 0.0 (middle panel), and $B/T$ = +0.4 (right panel). The point $(p_+^{(eq)}, p_-^{(eq)})$, computed using Eq. (S4), is marked by the red circle in each case; the point (1/3, 1/3) is marked by the green square. The distribution flows to the fixed point $(p_+^{(eq)}, p_-^{(eq)})$ for any value of $B/T$, independent of the initial conditions.

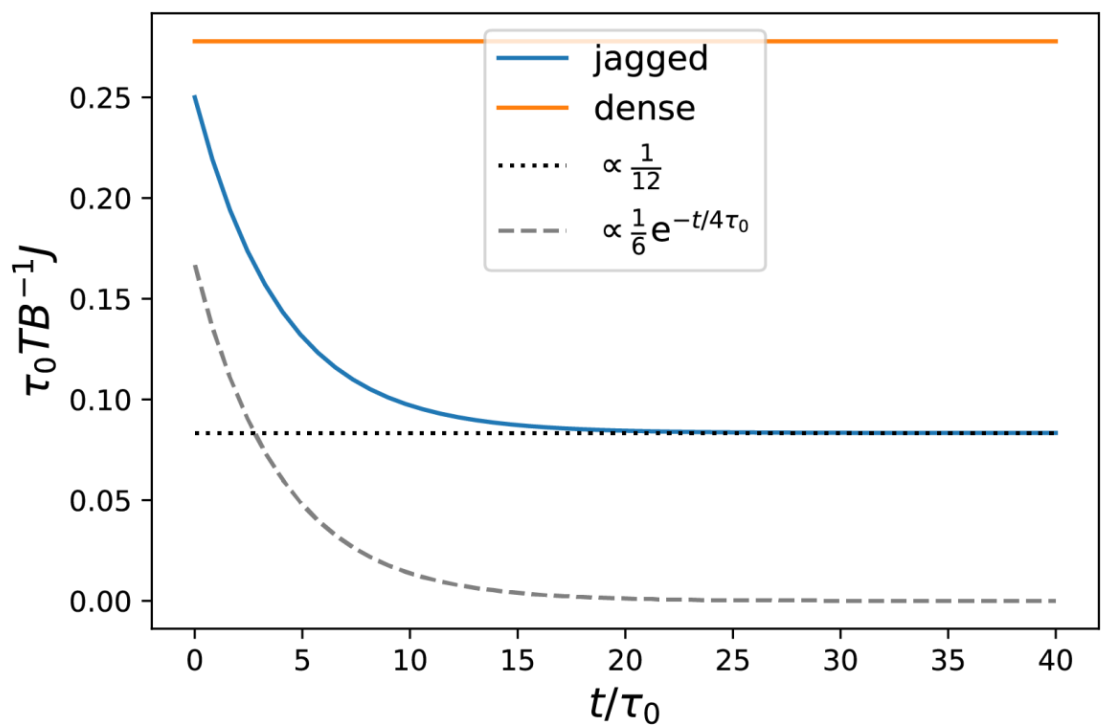

**Fig. S5**: The currents that arise on the two ladders when a small ($B/T \ll 1$) positive field is applied at time $t = 0$.

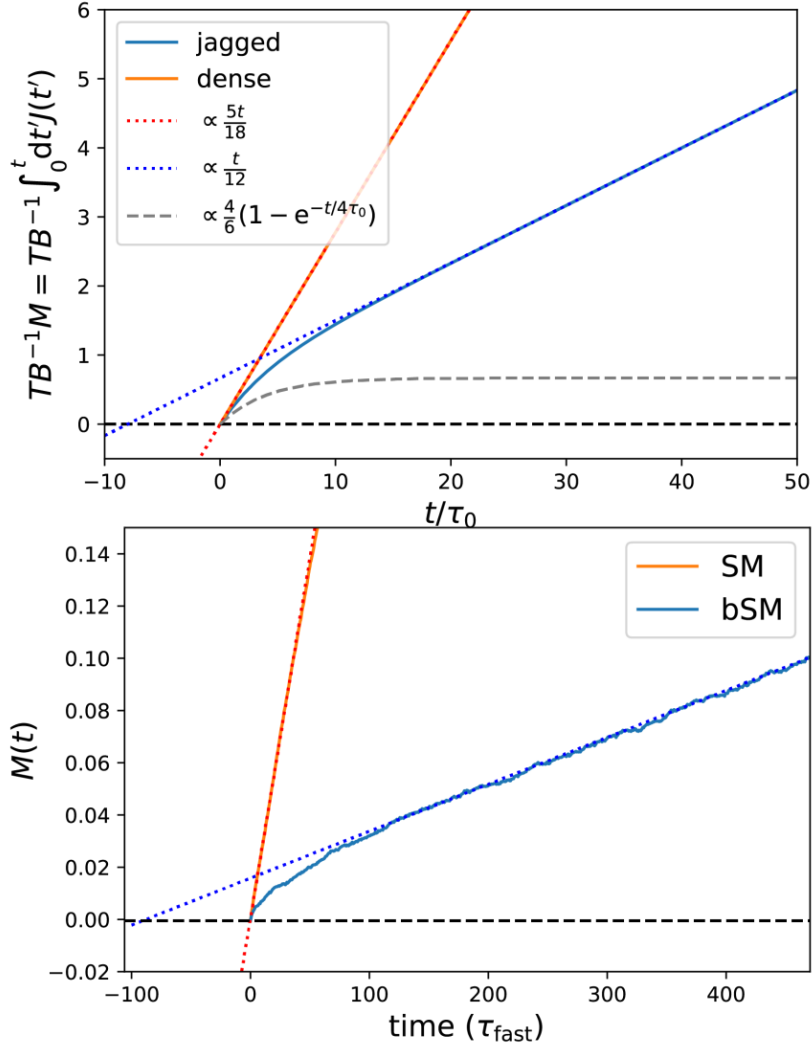

**Fig. S6**: At short times after a field is turned on, SM spin ice behaves similarly to the dense ladder and bSM spin ice behaves similarly to the jagged ladder. Top: The analogue of a magnetisation on the ladders, given by integrating Eq. (S19) and (S20). The straight dotted blue line, which is $\propto t/12$, intercepts the $M = 0$ line (dashed black) at $t \approx -10\tau_0$, and meets the $M(t)$ curve for the jagged ladder at $t \approx +10\tau_0$. Bottom: Example of the magnetisation change

*51*

in nearest-neighbour spin ice, for a system of 128,000 spins, after a field of magnitude 60 mT is turned on in the [100] direction, using SM and bSM dynamics and $\mathcal{H}_{NN}$ at $T/J_{eff}$ = 0.15 (corresponding to approximately 0.8 K). The straight dotted blue line intercepts the $M = 0$ line (dashed black) at $t \approx -100\tau_{fast}$, and meets the $M(t)$ curve for bSM at $t \approx +100\tau_{fast}$.

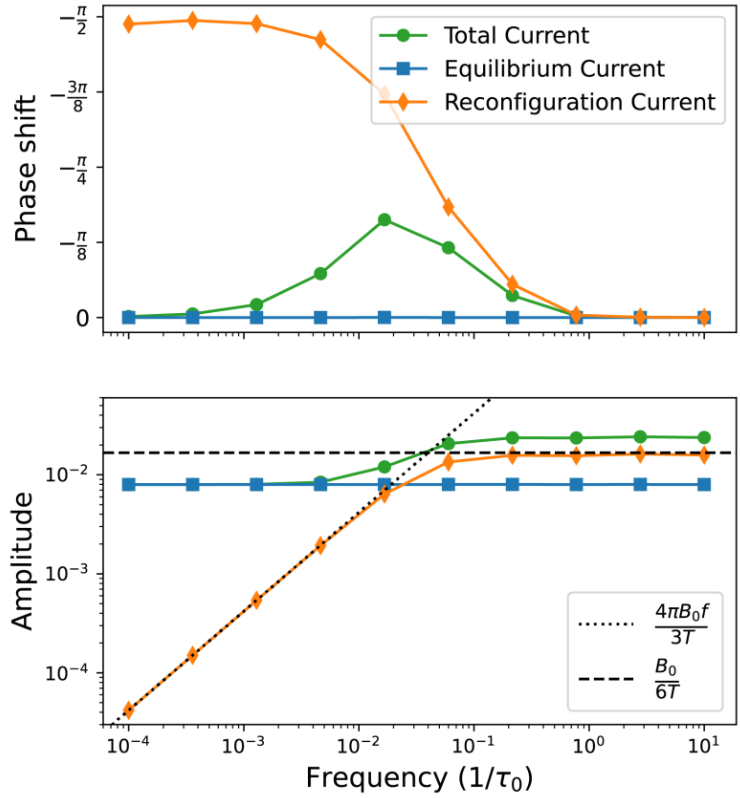

**Fig. S7**: Phase shift $\phi$ (top panel) and amplitude A (bottom panel) extracted from fits to the function $A\sin(2\pi ft - \phi)$, where $f$ is the frequency of the driving field, which has the form $B(t) = B_0\sin(2\pi ft)$. This was repeated for the currents $J_{jagged}$, $J_{ss}$, and $J_{rec}$. Examples of the probabilities and currents are given in Fig. S6. In the bottom panel, the dotted black line indicates the prediction for the amplitude if the probabilities $p_-$ and $p_+$ are in equilibrium

with the field, Eq. (S23), whereas the dashed black line indicates the predicted amplitude plateau at high frequency.

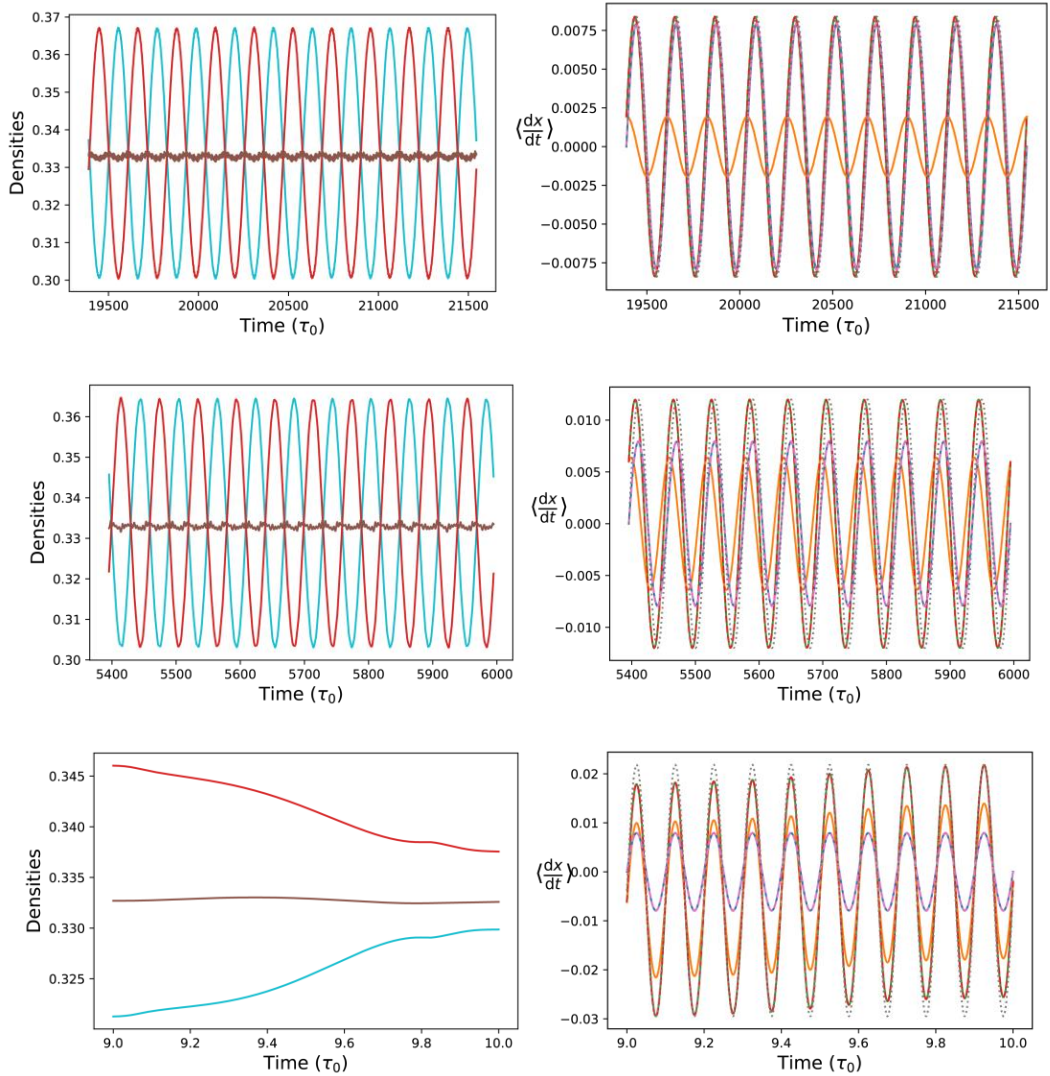

**Fig. S8**: Examples of the evolution of the probabilities (left) and currents (right) for a particle on a jagged ladder in an applied field oscillating (from top to bottom) at frequencies 0.005, 0.017, and 10.0 $\tau_0^{-1}$. The probabilities $p_0$ (brown), $p_-$ (cyan), and $p_+$ (red) evolve according to Eq. (S2) and Eq. (S3). From these, one can compute the total current (green) using Eq. (S5), the steady state current (blue) using Eq. (S6), and the reconfiguration current (orange) using Eq. (S7). The dotted grey lines indicate the applied field and the dashed red line in the right panels show the sum of $J_{ss}$ and $J_{rec}$, demonstrating that this matches the total current. All results were obtained by integrating numerically the differential equations governing the evolution of $p_+$ and $p_-$ (Eq. (S2) and (S3)), with initial condition $p_- = p_+ = 1/3$. The shown results correspond to the time window between 90 and 100 periods after the driving field is first turned on – after the system has settled into a steady state. The values $\tau_0 = 1$ and $B_0/T = 0.1$ were used in the figure.

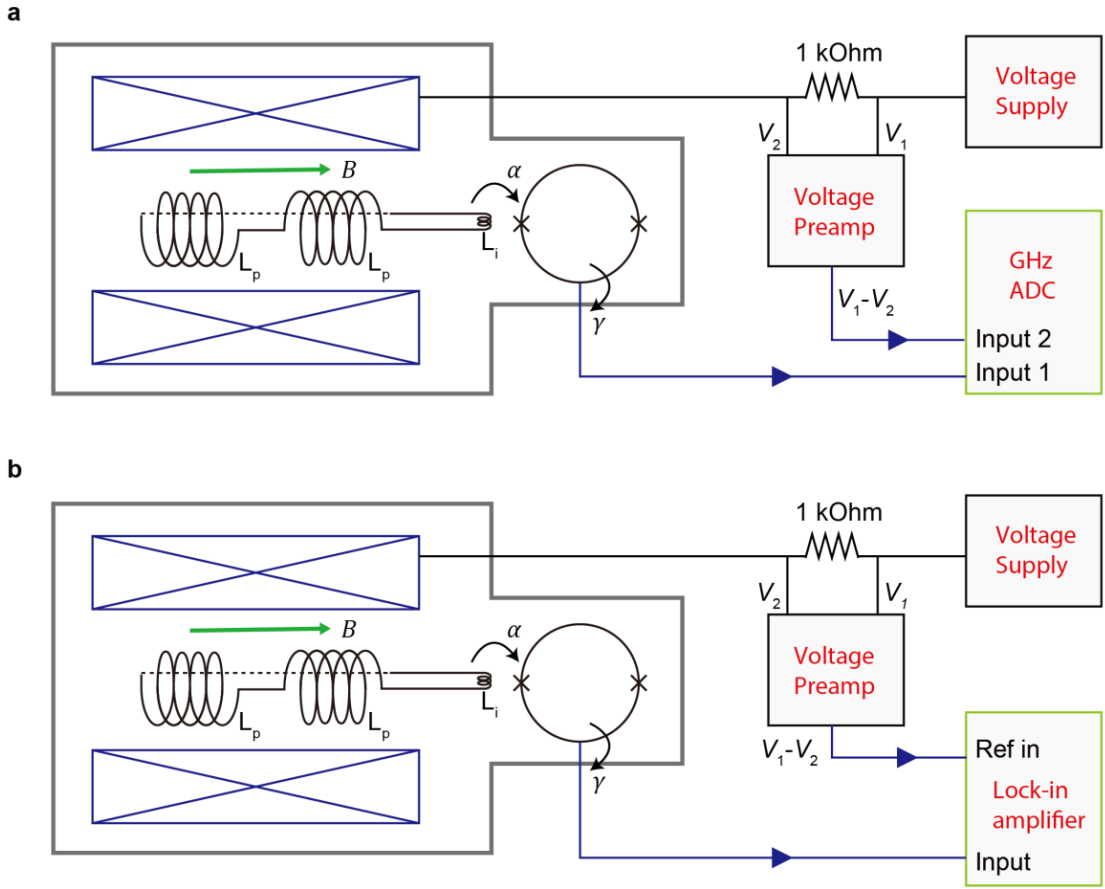

**Fig. S9 Schematic circuit diagram of the experiment. a**. Circuit diagram of the time-domain flux transient measurement. **b**. Circuit diagram of the frequency-domain conductivity measurement. The green box marks the main difference of electronics in two experiments.

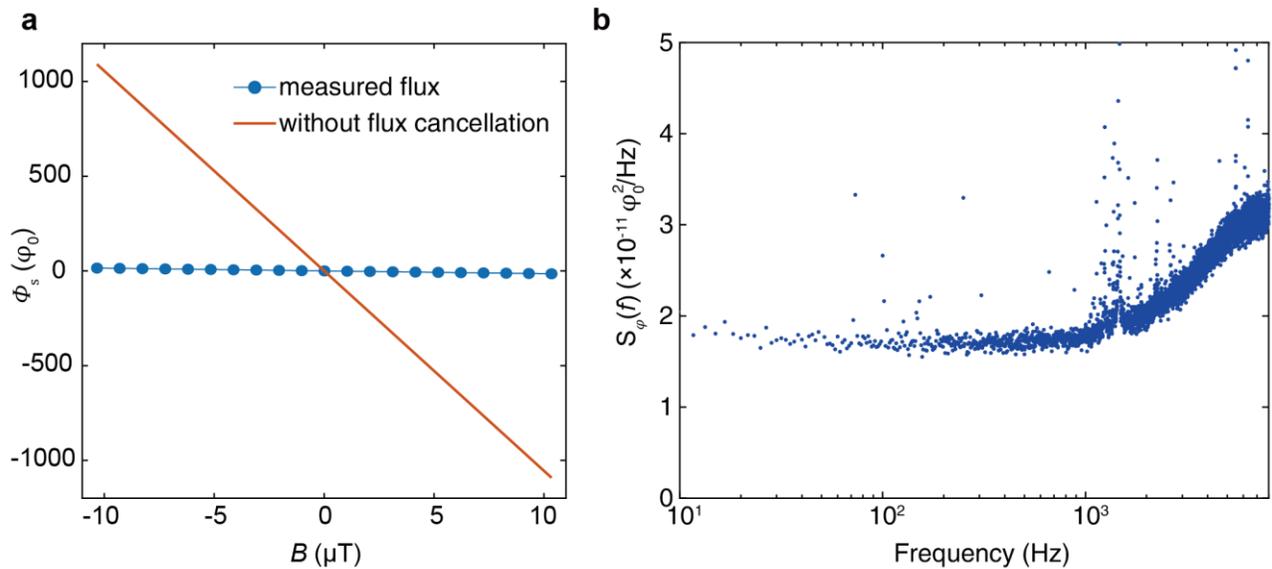

**Fig. S10 Performance of the compensation pickup coil. a.** Cancellation of external field with the counter-wound pickup coil. Here we show the measured flux $\Phi_S$ (blue line) for the empty coil calibration without sample at 4 K. The red line is the expected flux response if there is no compensation pickup coil. Only 1.3% of the residue flux is detected when the external field is applied. **b.** Noise performance of pickup coil under zero field at T = 4 K. The noise level is below $10^{-10}$ $\varphi_0^2$/Hz for the measured frequency.

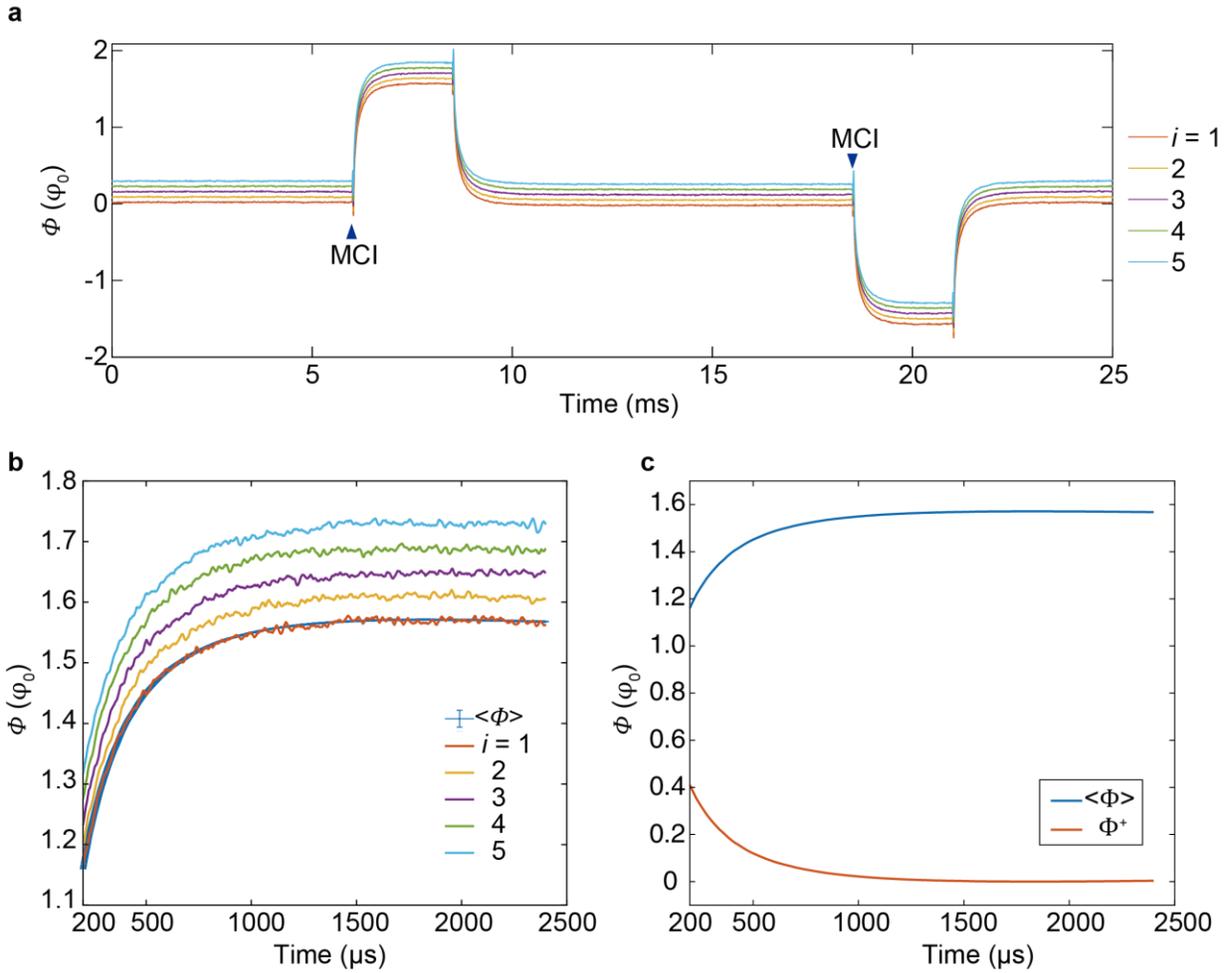

**Fig. S11 Schematic illustration of the averaging process. a**. Flux response measured by SQUID first cut into $k$ sections. The first five sections are shown as an example. Each flux line is shift for clarity. Typically, for each temperature over 1000 sections are averaged. **b**. The time window for the first 2500 µs after the magnetic field is switched on. The time at 0 µs in 2b is the at 6 ms (MCI) in 2a. The averaged flux response demonstrates significantly improved signal-to-noise. **c.** Comparison of averaged flux $\langle \Phi(t) \rangle$ and $\Phi^+$, where the $\Phi^+ = \langle \Phi(t = 1.8 \text{ ms}) \rangle - \langle \Phi(t) \rangle$.

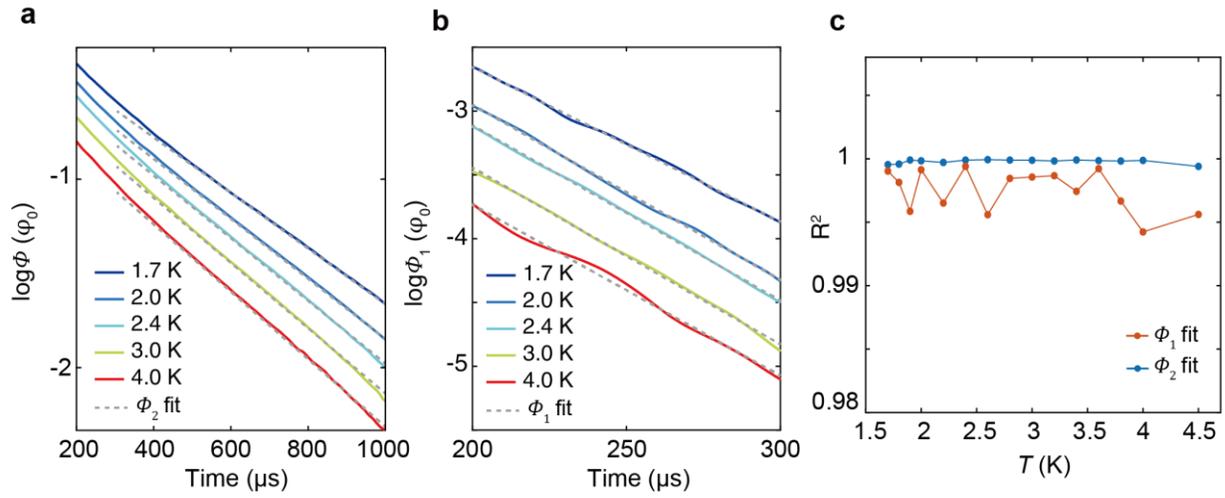

**Fig. S12 Demonstration of fit quality for monopole current dichotomy. a**. Linear fit of $\log \Phi_2(t, T)$ from the range of 600 μs to 1000 μs. The fitting curve is extended to 300 μs as a guidance for deviation. **b**. Linear fit of $\log \Phi_1(t, T)$ from 200 μs to 300 μs. **c.** $R^2$ of the linear fit of $\Phi_1$ and $\Phi_2$ for the measured temperature range, showing a consistent high fitting quality.

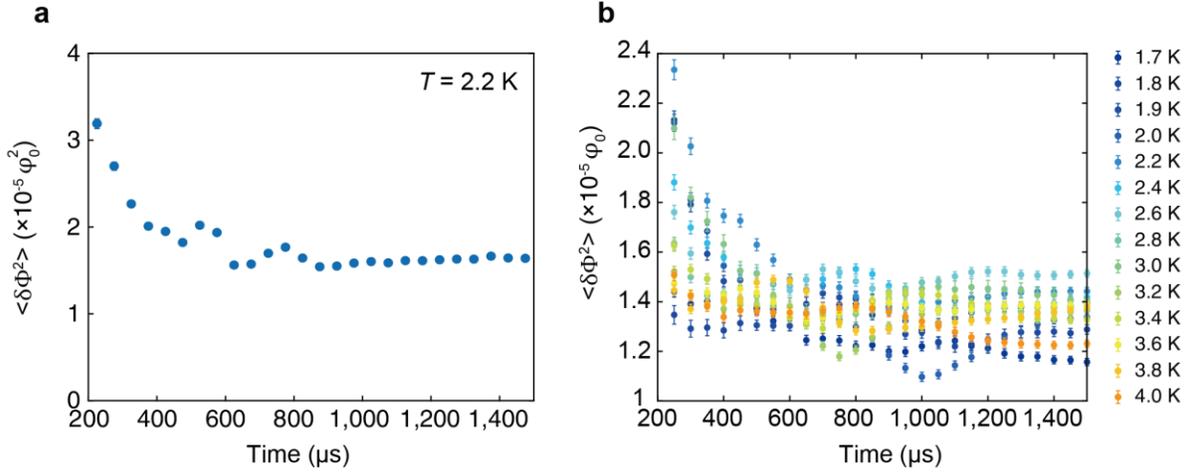

**Fig. S13 Time-dependent flux variance. a**. Time series flux variance at 2.2 K. Each variance point is an average of a bin within 50 μs, showing a sudden increase of flux noise at the first 100 μs. **b**. Temperature dependence of time series of flux variance. Each variance point is an average of a bin within 100 μs.

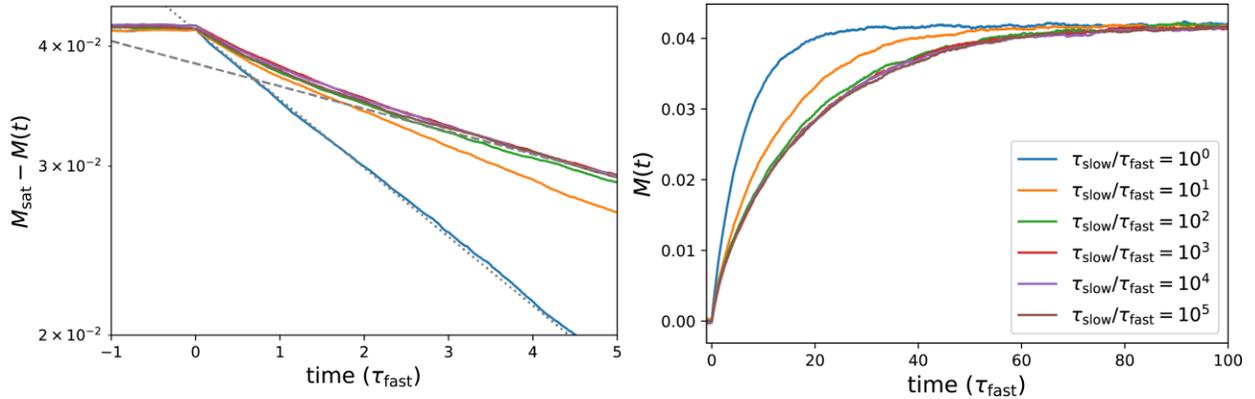

**Fig. S14**: The response to a sudden applied magnetic field of magnitude 15 mT for bSM dynamics with different values of $\tau_{slow}/\tau_{fast}$ at $T/J_{eff}$ = 0.3 (corresponding to approximately 1.4 K). These simulations used $\mathcal{H}_{NN}$ (Eq. (S1)) and a 128,000 spin system. The right panel shows the behaviour of the magnetisation $M(t)$, whereas the left panel shows $M_{sat} - M(t)$ in semi-logarithmic scale, highlighting the two regimes discussed in the main text. The dashed and dotted lines highlight the polarisation regime for large and small $\tau_{slow}/\tau_{fast}$, respectively.

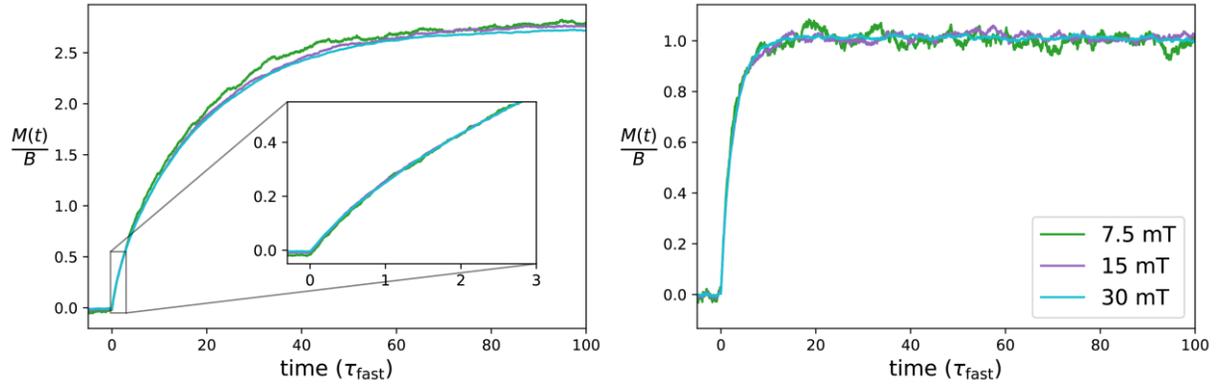

**Fig. S15**: The response of the magnetisation when a magnetic field is turned on for a system at $T/J_{eff}$ = 0.3 ($T \approx$ 1.4 K; left panel) and $T/J_{eff}$ = 0.7 ($T \approx$ 2.9 K; right panel). Three different field strengths were used, as indicated in the legend, and the magnetisation is normalised by the magnitude of the applied field. These simulations used $\mathcal{H}_{NN}$ and a 128,000 spin system.

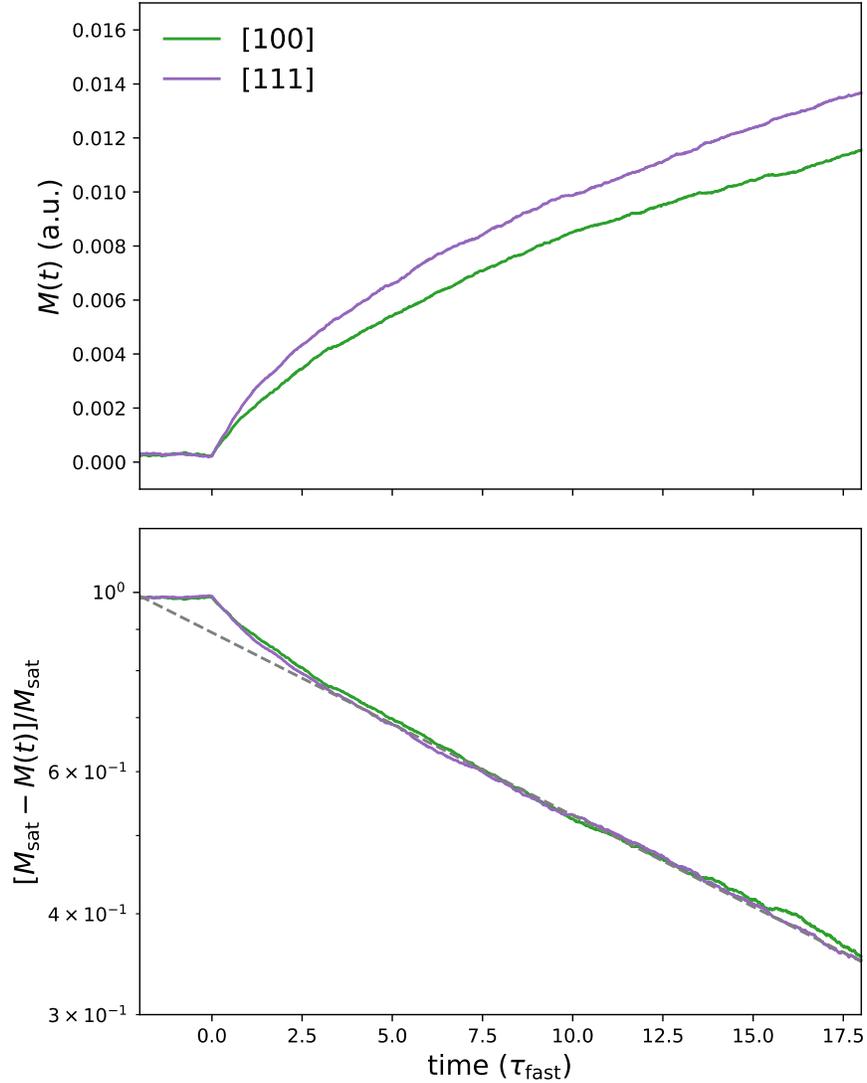

**Fig. S16**: The response of the magnetisation when a magnetic field of strength 7.5 mT is turned on for a system at $T/J_{\text{eff}}$ = 0.3 ($T \approx$ 1.4 K). The figure shows the response when the field is applied either along the [100] or the [111] lattice direction, with the magnetization measured along the same direction in each case. The response has the same form in both cases, but the magnitude of response is larger when the field is applied along the [111]

direction, as the field then couples more strongly to (one quarter of) the spins. In the bottom panel we show the response normalized by the magnetization at equilibrium with the applied field, $M_{\text{sat}}$, showing that the timescales and shape of the response are direction-independent. The dashed grey line is an exponential fit to the data for times $t > 2\tau_{\text{fast}}$. These simulations used $\mathcal{H}_{\text{NN}}$ and a 128,000 spin system.

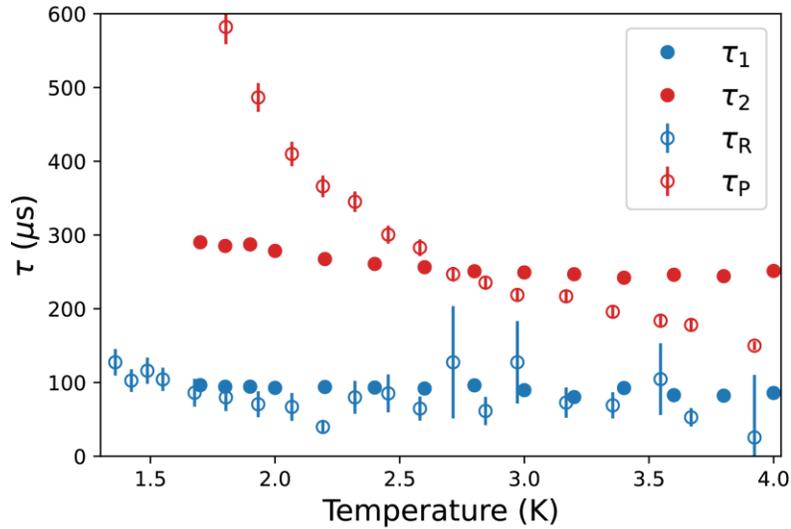

**Fig. S17**: The time scales $\tau_1$ (solid blue) and $\tau_2$ (solid red), governing the decay of the two currents in experiments, are compared to the time scales $\tau_R$ (unfilled blue) and $\tau_P$ (unfilled red), governing the decay of the reconfiguration and polarisation currents respectively in simulations. Simulations were performed with $\mathcal{H}_{\text{NN}}$ and a 128,000 spin system. Here $\tau_{\text{fast}}$ = 85 μs was used to convert Monte Carlo time into physical time.

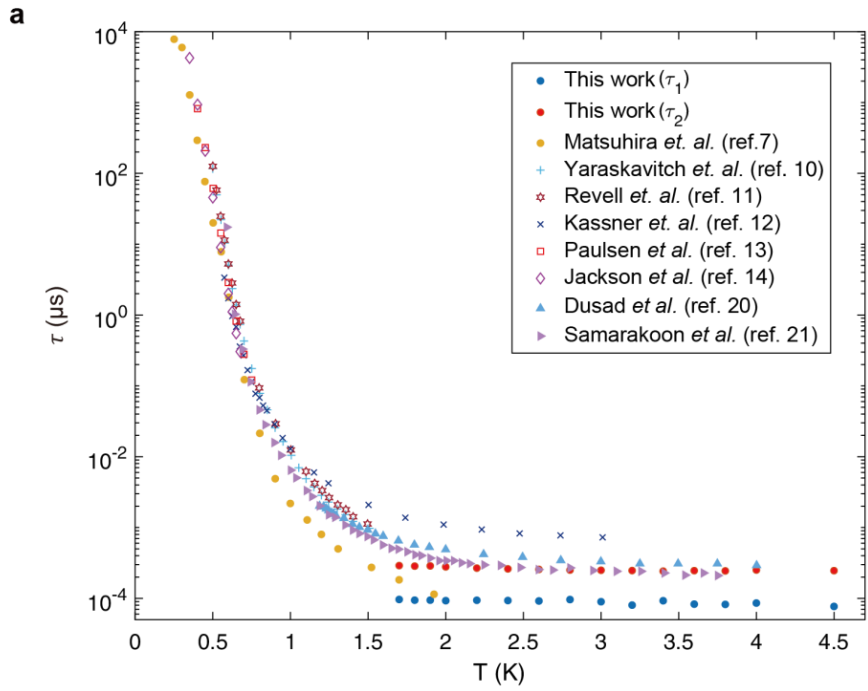

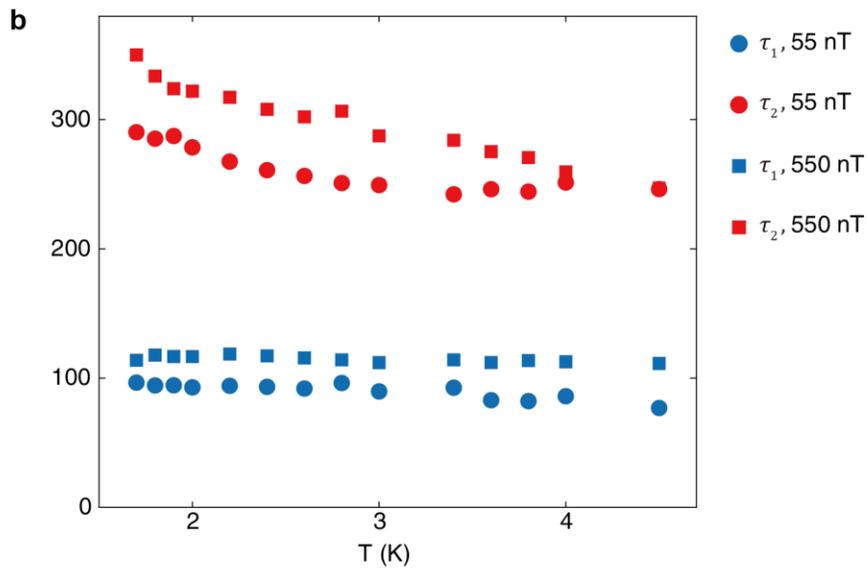

**Fig. S18**: **a.** Temperature dependence of the measured relaxation time constant. The fast and slow decaying time constant from this work are plotted with the relaxation time constants extracted from AC susceptibility data (ref. 7,10,12), correlation (ref. 11) and DC magnetization (ref. 13, 14). **b.** Temperature dependence of the extracted $\tau_1$ and $\tau_2$ from the time-domain measurement at two different magnetic field (55 nT & 550 nT).

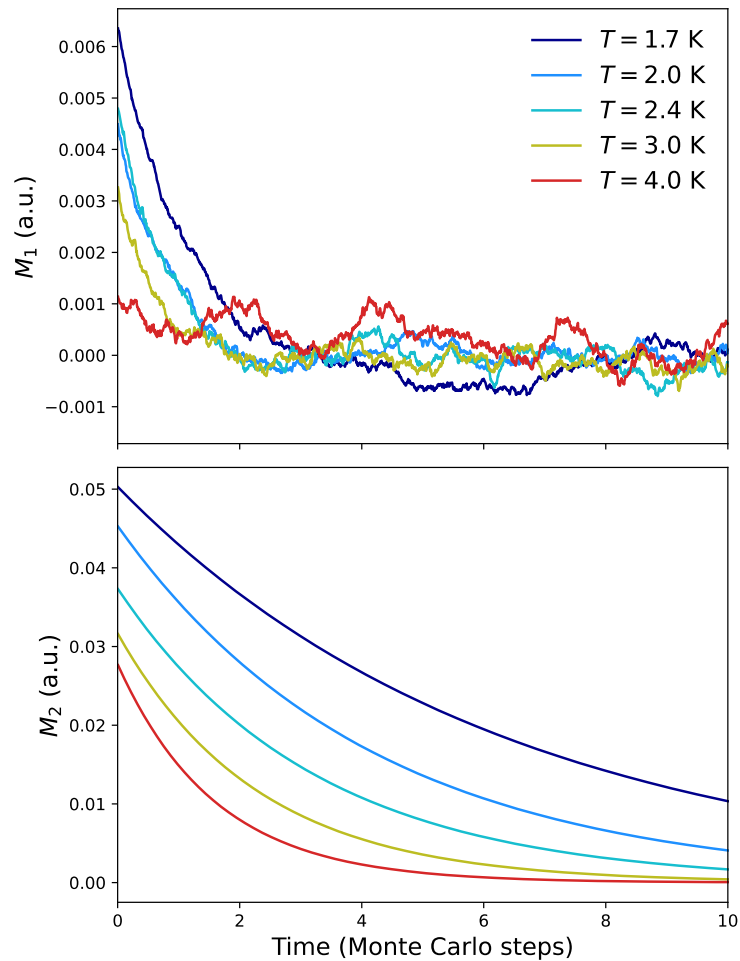

**Fig. S19:** Extracted fast and slow magnetization response from Monte Carlo simulations

when a field of strength 30 mT is suddenly applied at $t = 0$. The data shown here are the same as those shown in Fig. 1B in the main text, and were measured using $\mathcal{H}_{\text{OP}}$ (Eq. S1) and a 16,000 spin system. The plotted contributions are defined as $M_1 = M_{\text{sat}} - M - A_2 e^{-t/\tau_2}$ and $M_2 = A_2 e^{-t/\tau_2}$, where $A_2$ and $\tau_2$ are parameters extracted by fitting the equation for $M_2$ to $M_{\text{sat}} - M$ for times $t > 1.6$ Monte Carlo steps. A single relaxation time cannot account for the observed data for temperatures up to at least 3 K. $M_1$ does not show any clear decay at $T = 4$ K. However, at this high temperature the extraction of $M_1$ suffers from numerical noise limitations, and we are hesitant to draw any conclusions about the existence or non-existence of two relaxation time scales at 4 K from these data.

## Supplementary References